\documentclass[%
reprint,
superscriptaddress,
 amsmath,amssymb,
 aps,
]{revtex4-1}

\usepackage{graphicx}
\usepackage{dcolumn}
\usepackage{bm}
\usepackage[mathlines]{lineno}
\usepackage[colorlinks=true]{hyperref}
\usepackage{mathrsfs}
\usepackage{xcolor}
\usepackage{float}
\usepackage{ulem}
\usepackage[utf8]{inputenc}

\newcommand{\be}{\begin{equation}}
\newcommand{\ee}{\end{equation}}
\newcommand{\bea}{\setlength\arraycolsep{2pt} \begin{eqnarray}}
\newcommand{\eea}{\end{eqnarray}}

\newcommand{\bal}{\begin{aligned}}
\newcommand{\eal}{\end{aligned}}

\begin{document}


\title{Can shadows reflect phase structures of black holes?}
\author{Ming Zhang}
\email{mingzhang@jxnu.edu.cn}
\affiliation{Department of Physics, Jiangxi Normal University, Nanchang 330022, China}
\author{Minyong Guo}
\email{Corresponding author: minyongguo@pku.edu.cn}
\affiliation{Department of Physics and State Key Laboratory of Nuclear Physics and Technology, Peking University, No.5 Yiheyuan Rd, Beijing 100871, China}
\affiliation{Center for High Energy Physics, Peking University, No.5 Yiheyuan Rd, Beijing 100871, China}

\date{\today}

\begin{abstract}
The relation between the black hole shadow and the black hole thermodynamics is investigated. We find that the phase structure can be reflected by the shadow radius for the spherically symmetric black hole. We also find that the shadow size gives correct information but the distortion of the shadow gives wrong information of the phase structure for the axially symmetric black hole.
\begin{description}
\item[Key words]
Black Hole Shadow, Black Hole Thermodynamics, Phase Structure
\end{description}
\end{abstract}


\maketitle


\section{Introduction}
Recently, the event horizon telescope (EHT) has given the first image of the supermassive M87${}^{*}$ black hole in the central giant elliptical galaxy \cite{Akiyama:2019cqa,Akiyama:2019brx,Akiyama:2019sww,Akiyama:2019bqs,Akiyama:2019fyp,Akiyama:2019eap}, which gave direct support of the Einstein's general relativity and the existence of the black hole in our universe. The image of the black hole can provide us with information about jets and matter accretion of the black hole. Moreover, the shadow of a black hole can tell us information about the black hole, such as the mass and the rotation parameters of the black hole.

The shadow of the black hole, due to the gravitational lensing of light, is defined as the observer's dark sky, where the light sources are distributed everywhere except the region between the observer and the black hole. The shadow of a spherically symmetric black hole is a black circular disk. The shadow of a Schwarzschild black hole, which is dependent on the black hole mass and the position of the observer, was first investigated in \cite{synge1966escape,luminet1979image}. The shadow of a rotating Kerr black hole, which is elongated silhouette-like due to the rotation's dragging effect, was first worked out in \cite{hawking1973black}. The shadow of the Kerr-Newman black hole was studied in \cite{de2000apparent,Hioki:2008zw}. For the whole class of Plebański-Demiański spacetime, the shadow was studied in \cite{Grenzebach:2014fha}. The shadows of various other black holes have been studied, {\it e.g.}, see Refs. \cite{Wang:2017qhh,Guo:2018kis,Yan:2019etp,Hennigar:2018hza,Konoplya:2019sns,Bambi:2008jg,Amir:2017slq,Bambi:2010hf,Konoplya:2019fpy,Jusufi:2020cpn,Wang:2019tjc,Zeng:2020dco,Wei:2019pjf}. Moreover, the shadows of the wormholes are investigated \cite{Ohgami:2015nra,Nedkova:2013msa,Shaikh:2018kfv,Amir:2018szm,Amir:2018pcu}. The shadow perturbed by the gravitational waves was studied in \cite{Wang:2019skw}. For a brief review, one can see \cite{Cunha:2018acu}.

Black hole thermodynamics has been widely researched,  {\it e.g.}, see Refs. \cite{Kubiznak:2012wp,Wei:2019uqg,Cai:2013qga,Zhang:2018djl}. Black holes are viewed as thermodynamic systems, which own thermodynamic properties that are quite similar to those of ordinary objects. There are four thermodynamic laws for the black holes \cite{Bardeen:1973gs}. The first law can be written as
\begin{equation}
dU=TdS+X_{i}dY_{i},
\end{equation}
where $U$ is the internal energy, $S$ is the Bekenstein-Hawking entropy, $X_{i}$ are the generalized displacements, and $Y_{i}$ are the generalized forces. The second law claims that the event horizon area of the black hole will not decrease with time, {\it{i.e.}},
\begin{equation}
\delta A\geqslant 0.
\end{equation}
The third law says that the surface gravity $\kappa$ cannot be reduced to zero by finite operations. The zeroth law reads the surface gravity is constant over the event horizon for stationary black holes. 

One can see that all the four thermodynamic laws for the black hole are related to the event horizon, a null hyper-surface that holds the symmetric characteristics of the spacetime. The event horizon plays a key role in  reflecting the physical properties of the black hole. Notwithstanding, some endeavours are trying to replace the role of event horizon with the photon orbit around the black hole. Based on the case studies in \cite{Wei:2018aqm,Han:2018ooi}, it was found that the circular orbit radius of the photon can be a characteristic quantity to reflect the thermodynamic phase structure of a spherically symmetric black hole \cite{Zhang:2019tzi}. Related extensions can also be seen in \cite{Xu:2019yub,Li:2019dai}. The image of the black hole makes us think about whether it is possible to use the black hole shadow, which is observable, to reflect the thermodynamics of the black hole. In this compact paper, we will introduce our thoughts on this. Besides the event horizon and circular orbit radius of the photon, we try to use the third quantity, the radius of the shadow, to investigate the thermodynamic information of the black holes, both for spherically symmetric one and axially symmetric one. The motivation of using the radius of the black hole shadow as a characteristic quantity to detect the thermodynamics of the black hole is that this new quantity can be observed, such that thermodynamics of the black hole may be tested by observable data. 

In the following section, we will individually study the relations between the shadow sizes and phase structures for the general spherically symmetric black hole and axially symmetric black holes (using the Kerr  black hole with a cosmological constant  as a toy model). We will give our conclusion in section \ref{remarks}.

\section{Shadows and thermodynamic phase structures of black holes}\label{satp}
The specific heat of a black hole, according to the first law of the black hole thermodynamics, can be written as
\begin{equation}
C=\frac{dU}{dT}=T\left(\frac{\partial S}{\partial T}\right)_{Y_i},
\end{equation}
$C>0$, $C<0$ correspond to the thermodynamically unstable state and stable state individually. The point making $C=0$ is the phase transition point. In general, the entropy $S$ is related to the event horizon $r_h$ of the black hole with
\begin{equation}
\frac{dS}{dr_h}>0.
\end{equation}
Accordingly, we can know that 
\begin{equation}\label{sig}
\text{Sgn}(C)= \text{Sgn}\left(\frac{\partial T}{\partial r_h}\right),
\end{equation}
where Sgn is the sign function. Thus, in the $r_h -T$ diagram, a positive slope means that the black hole is in a thermodynamically stable state and a negative slope corresponds to a thermodynamically unstable black hole. In the following, we will show that whether the shadow radii of the black holes can be used  to mirror the phase structures of the black holes or not, first for the spherically symmetric case and then for the axially symmetric case.

\subsection{Shadow and phase structure of the spherically symmetric black hole}

The static spherically symmetric spacetime background can be described by the line element
\begin{equation}\nonumber
ds^{2}=-f(r)dt^{2}+\frac{dr^{2}}{g(r)}+r^{2}d\theta^{2}+r^{2}\sin^{2}\theta d\phi^{2},
\end{equation}
where $f(r), g(r)$ are functions of the coordinate $r$. The Hamiltonian of the photon moving in the static spherically symmetric spacetime is
\begin{equation}\label{hamij}
2H=g^{ij}p_{i}p_{j}=0.
\end{equation}
Without loss of generality, due to the spherically symmetric characteristics of the black hole, we can consider photons moving in the equatorial plane with $\theta=\pi/2$. Then (\ref{hamij}) can be explicitly written as
\begin{equation}\label{ramo}
\frac{1}{2}\left[-\frac{p_{t}^{2}}{f(r)}+g(r)p_{r}^{2}+\frac{p^{2}_{\phi}}{r^{2}}\right]=0.
\end{equation}
Thinking of the relations
\begin{equation}
\dot{p}_{t}=-\frac{\partial H}{\partial t}=0
\end{equation}
and
\begin{equation}
\dot{p}_{\phi}=-\frac{\partial H}{\partial \phi}=0,
\end{equation}
we can know that $p_{t}$ and $p_{\phi}$ are constants of motion. We can define $-p_{t}=e$ and $p_{\phi}=j$, which can be physically interpreted as the energy and angular momentum of the photon.

The equations of motion for the photon can be obtained as
\begin{equation}
\dot{t}=\frac{\partial H}{\partial p_{t}}=-\frac{p_{t}}{f(r)},
\end{equation}
\begin{equation}
\dot{\phi}=\frac{\partial H}{\partial p_{\phi}}=\frac{p_{\phi}}{r^{2}},
\end{equation}
\begin{equation}
\dot{r}=\frac{\partial H}{\partial p_{r}}=p_{r}g(r),
\end{equation}
where $p_{r}$ is the radial momentum. The effective potential of the photon can be defined by
\begin{equation}
V_{e}+\dot{r}^{2}=0.
\end{equation}
As a result, the effective potential can be expressed as
\begin{equation}\label{e1}
V_{e}=g(r)\left[\frac{j^{2}}{r^{2}}-\frac{e^{2}}{f(r)}\right].
\end{equation}
The radius of the photon can be obtained from the relation
\begin{equation}\label{effpho}
V_{e}=V^{\prime}_{e}=0.
\end{equation}

The orbit equation for the photon is
\begin{equation}\label{oreq}
\frac{dr}{d\phi}=\frac{\dot{r}}{\dot{\phi}}=\frac{r^{2}g(r)p_{r}}{j}.
\end{equation}
Substituting the solution of $p_{r}$ from (\ref{ramo}) into (\ref{oreq}), we can obtain
\begin{equation}
\frac{dr}{d\phi}=\pm r\sqrt{g(r)\left[\frac{r^{2}e^{2}}{f(r)j^{2}}-1\right]}.
\end{equation}
At the turning point of the photon orbit, we should have
\begin{equation}
\left.\frac{dr}{d\phi}\right|_{r=R}=0,
\end{equation}
which gives
\begin{equation}
\frac{e^{2}}{j^{2}}=\frac{f(R)}{R^{2}}.
\end{equation}
Then we can know
\begin{equation}\label{eosf}
\frac{dr}{d\phi}=\pm r\sqrt{g(r)\left[\frac{r^{2}f(R)}{f(r)R^{2}}-1\right]}.
\end{equation}

For a light ray sending from a static observer at position $r_{o}$ and transmitting into the past with an angle $\alpha$ with respect to the radial direction, we have 
\begin{equation}
\cot\alpha=\frac{\sqrt{g_{rr}}}{\sqrt{g_{\phi\phi}}}\cdot\frac{dr}{d\phi}{\Big{|}}_{r=r_{o}}=\frac{1}{r\sqrt{g(r)}}\cdot\frac{dr}{d\phi}{\Big{|}}_{r=r_{o}}.
\end{equation}
Thinking of (\ref{eosf}), we can have 
\begin{equation}
\cot^{2}\alpha=\frac{r_{o}^{2}f(R)}{f(r_{o})R^{2}}-1.
\end{equation}
Using the relation $1+\cot^{2}\alpha=\sin^{-2}\alpha$, we further have
\begin{equation}
\sin^{2}\alpha=\frac{f(r_{o})R^{2}}{r_{o}^{2}f(R)}.
\end{equation}
The angular radius of the black hole shadow can be obtained by letting $R\to r_{p}$ with $r_{p}$ the circular orbit radius of the photon. We can obtain the shadow radius of the black hole observed by a static observer at position $r_{o}$ as
\begin{equation}\label{shara}
r_{s}=r_{o}\sin\alpha=\left. R\sqrt{\frac{f(r_{o})}{f(R)}}\right|_{R\to r_{p}}.
\end{equation}

\begin{figure}[!htbp] 
   \centering 
   \includegraphics[width=2.0in]{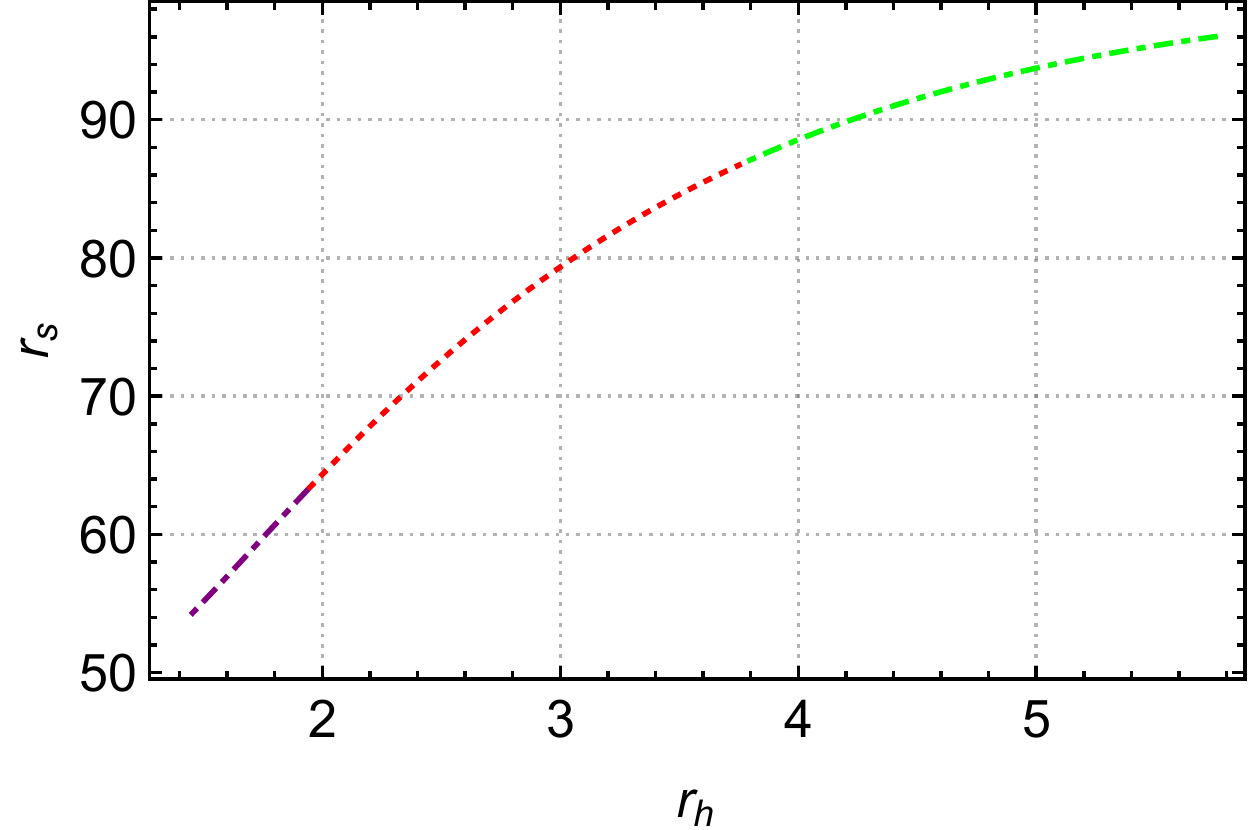}\\
   \includegraphics[width=2.25in]{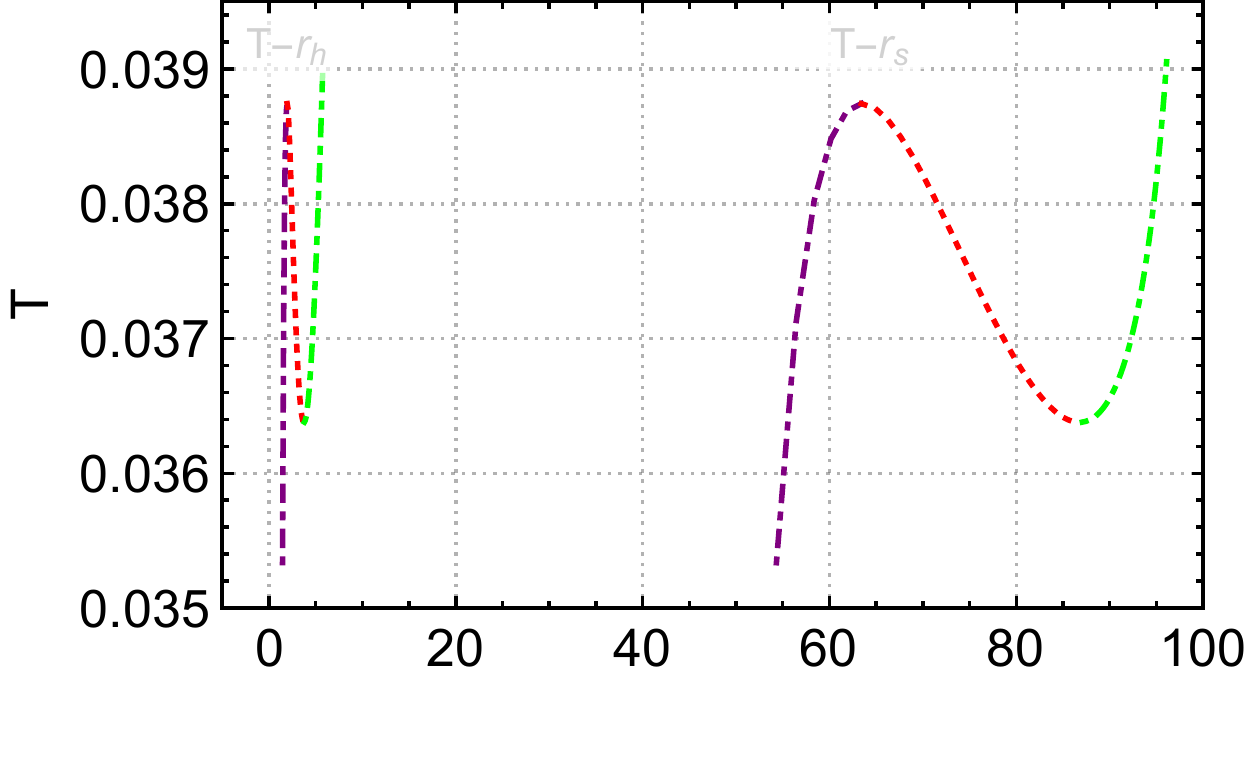}\,\,\,
   \caption{The upper diagram shows the relation between the shadow radius and the event horizon for the RN-AdS black hole. The bottom diagram shows the temperature in terms of the event horizon and the shadow radius for the black hole, respectively. We set the charge of the black hole as 1 and the cosmological constant as $-4\pi/225$.  The purple, red and green curves in the top diagram correspond to the curves with the same colours in the bottom diagram.}
   \label{figx1}
\end{figure}

For the circular orbit of the photon, we have 
\begin{equation}
V^{\prime\prime}_{e}<0,
\end{equation}
combing it with (\ref{e1}) and (\ref{effpho}), we have
\begin{equation}
\begin{aligned}
\frac{j^2 g(r) \left[r f'(r)-2 f(r)\right]}{r^3 f(r)}&<0, \quad  \text{for}\quad r>r_p;\\ \frac{j^2 g(r) \left[r f'(r)-2 f(r)\right]}{r^3 f(r)}&>0, \quad  \text{for}\quad r<r_p.
\end{aligned}
\end{equation}
Then using (\ref{shara}), we can further get
\begin{equation}\label{dre1}
\left.\frac{dr_s}{dR}\right|_{R\to r_p}=\left[\frac{f\left(r_0\right)}{f(R)}\right]^{3/2}
\left.\frac{2 f\left(R\right)-r_p f'\left(R\right)}{2 f\left(r_0\right)}\right|_{R\to r_p}>0
\end{equation}
for $R-r_p >0^+$. Additionally,  (\ref{e1}) together with (\ref{effpho}) gives
\begin{equation}
f(r_p)=\frac{e^2}{j^2}r_p^2
\end{equation}
and
\begin{equation}
\left.\frac{df(R)}{dR}\right|_{R\to r_p}>0.
\end{equation}
As the Hawking temperature of the black hole is positive, we know
\begin{equation}
\left.\frac{df(r)}{dr}\right|_{r\to r_h}=\left.\frac{df(R)}{dR}\frac{dR}{dr_h}\right|_{R\to r_p}>0,
\end{equation}
which implies
\begin{equation}\label{dre2}
\left.\frac{dR}{dr_h}\right|_{R\to r_p}>0.
\end{equation}
Considering (\ref{dre1}) and (\ref{dre2}), we know
\begin{equation}\label{dre3}
\frac{dr_s}{dr_h}>0
\end{equation}
for $R\to r_p$.
As a result, for the temperature $T$ of the black hole, we have 
\begin{equation}
\frac{\partial T}{\partial r_{h}}=\frac{\partial T}{\partial r_{s}}\frac{dr_{s}}{dr_{h}},
\end{equation}
which means
\begin{equation}\label{sphereone}
\frac{\partial T}{\partial r_{h}}>0,~~\frac{\partial T}{\partial r_{h}}=0,~~\frac{\partial T}{\partial r_{h}}<0
\end{equation}
correspond to
\begin{equation}\label{spheretwo}
\frac{\partial T}{\partial r_{s}}>0,~~\frac{\partial T}{\partial r_{s}}=0,~~\frac{\partial T}{\partial r_{s}}<0,
\end{equation}
respectively. That is, to be reminiscent of (\ref{sig}), the radius of the black hole shadow can also be a characteristic quantity to reflect the phase structure of the spherically symmetric black hole. One should note that $R\to r_{p}$ here does not mean that $R=r_{p}$; it means $R=r_{p}+\mathcal{O}(\epsilon)$ with $0\sim\epsilon\ll 1$, as we have $V^{\prime}_{e}=0$. So, exactly speaking, it is a position which is infinitely close to the outer edge of the shadow that conveys the thermodynamic information for the spherically symmetric black hole. In Fig. \ref{figx1}, we use the Reissner-Nordström-Anti-de Sitter (RN-AdS) black hole as a model to show that the shadow radius is indeed a fine quantity reflecting the phase structure of the spherically symmetric black hole. For simplicity, we here follow the conventions in Ref. \cite{Kubiznak:2012wp} and do not show specific expressions about the RN-AdS black hole.

\subsection{Shadow and phase structure of the axially symmetric black hole}
Here we will use the Kerr black hole with a cosmological constant (the well-known Kerr-(A)dS black hole) to show that the distortion and size of the shadow cannot be characteristic quantities to reflect the phase structure of the axially symmetric black hole. The Kerr black hole with a cosmological constant can be described by the line element using Boyer-Lindquist coordinates as
\begin{equation}
\begin{aligned}
ds^2=&-\frac{ \left(\Delta _r-a^2 \Delta _{\theta } \sin ^2\theta \right)dt^2}{\Sigma }+\Sigma  \left(\frac{d\theta^2}{\Delta _{\theta }}+\frac{dr^2}{\Delta _r}\right)\\ &+\frac{ \left[\Delta _{\theta } \sin ^2\theta  (a \chi +\Sigma )^2-\chi ^2 \Delta _r\right] d\phi^2}{\Sigma }\\&+\frac{2  \left[\chi  \Delta _r-a \Delta _{\theta } \sin ^2\theta (a \chi +\Sigma )\right] dt d\phi}{\Sigma },
 \end{aligned}
\end{equation}
where
\be\bal
\chi&=a\sin ^2\theta,\\
\Sigma&=r^{2}+a^2 \cos ^2\theta ,\\
\Delta_r&=r^2 \left[1-\frac{1}{3} a^2 \Lambda \right]+a^2-\frac{1}{3} r^4 \Lambda -2 m r,\\
\Delta_\theta&=1+\frac{1}{3} a^2 \Lambda \cos ^2\theta.
\eal\ee
$m$ is the mass parameter of the black hole, $a$ is the rotation parameter defined by $a\equiv J/M$ with $J$ the angular momentum of the black hole, and $\Lambda$ is the cosmological constant. The event horizon $r_{h}$ of the Kerr-AdS black hole or the Kerr black hole is the biggest real root of the equation $\Delta_{r}=0$ where we can also obtain the event horizon of the Kerr-dS black hole. The event horizon temperature of the Kerr-(A)dS black hole is
\begin{equation}\label{tempkerr}
\begin{aligned}
T=\frac{r_h \left(-\frac{a^2}{r_h^2}-\frac{a^2 \Lambda }{3}-\Lambda  r_h^2+1\right)}{4 \pi  \left(a^2+r_h^2\right)}.
\end{aligned}
\end{equation}
When $\Lambda=0$, it reduces to the temperature of the Kerr black hole. In Appendix \ref{ap}, we briefly review the thermodynamic quantities that will be used to analyze the phase structures of the black holes in what follows.

The null geodesics around the Kerr-(A)dS black hole is \cite{Grenzebach:2014fha,Li:2020drn}
\begin{subequations}
\begin{equation}
\frac{\Sigma}{\Omega^{2}}\frac{dt}{d\lambda}= \frac{\chi(l-E\chi)}{\Delta_{\theta} \sin^{2}\theta}
+\frac{(\Sigma + a\chi) \bigl((\Sigma + a\chi)E - al\bigr)}{\Delta_{r}},
\end{equation}
\begin{equation}
\frac{\Sigma}{\Omega^{2}} \frac{d\phi}{d\lambda}= \frac{l-E\chi}{\Delta_{\theta} \sin^{2}\theta}
		+\frac{a\bigl((\Sigma + a\chi)E - al\bigr)}{\Delta_{r}},
\end{equation}
\begin{equation}
\biggl(\frac{\Sigma}{\Omega^{2}}\biggr)^{2}\left(\frac{d\theta}{d\lambda}\right)^{2} =\Delta_{\theta}K - \frac{(\chi E - l)^{2}}{\sin^{2}\theta} =: \Theta(\theta),\label{loeff}
\end{equation}
\begin{equation}
\biggl(\frac{\Sigma}{\Omega^{2}}\biggr)^{2} \left(\frac{dr}{d\lambda}\right)^{2} = \bigl((\Sigma + a\chi)E-al\bigr)^{2} - \Delta_{r}K=: R(r),\label{reff}
\end{equation}
\end{subequations}
with $\lambda$  the affine parameter, $l$ the angular momentum of the photon, $E$ the energy of the photon (we here use different denotations from the spherically symmetric case),  $K$ the Carter constant relating with the Killing-Yano tensor field in the Kerr-(A)dS spacetime. $\Theta(\theta)$ and $R(r)$  are effective potential of the photon in the longitudinal and radial directions, respectively.

The radius of the circular orbit $r_{p}$ for the photon can be obtained by solving $R(r)=0$ and $dR(r)/dr=0$, which give
\be
K_{E} = \frac{16r^{2}\Delta_{r}}{(\Delta_{r}')^{2}}, \quad
L_{E} =\frac{ \Sigma + a\chi}{a} - \frac{4r\Delta_{r}}{a\Delta_{r}’},
\ee
where the $'$ means the derivative with respect to the radial coordinate $r$ and we have defined
\be
K_{E} \equiv \frac{K}{E^{2}}, \quad L_{E} \equiv \frac{l}{E}.
\ee

To obtain the contour of the black hole shadow, we can choose the zero-angular-momentum-observer (ZAMO) reference frame \cite{Stuchlik:2018qyz}
\begin{subequations}
\begin{eqnarray}
  \hat{e}_{(t)} & = & \sqrt{\frac{g_{\phi \phi}}{g_{t \phi}^2 - g_{t t} g_{\phi
  \phi}}} \left( \partial_t - \frac{g_{t \phi}}{g_{\phi \phi}} \partial_{\phi}
  \right),\label{et}\\
  \hat{e}_{(r)} & = & \frac{1}{\sqrt{g_{r r}}} \partial_r,\\
  \hat{e}_{(\theta)} & = & \frac{1}{\sqrt{g_{\theta \theta}}} \partial_{\theta},\\
  \hat{e}_{(\phi)} & = & \frac{1}{\sqrt{g_{\phi \phi}}} \partial_{\phi}.
\end{eqnarray}
\end{subequations}
Then the observation angles $(\alpha, \beta)$ can be defined by \cite{Cunha:2016bpi}
\begin{subequations}
\begin{eqnarray}
  p^{(r)} & = & p^{(t)} \cos \alpha \cos \beta, \\
  p^{(\theta)} & = &p^{(t)} \sin \alpha, \\
  p^{(\phi)} & = &p^{(t)} \cos \alpha \sin \beta,
\end{eqnarray}
\end{subequations}
where we have used the projections
\begin{eqnarray}
  p^{(t)} & = & - p_{\mu} \hat{e}_{(t)}^{\mu}, \\
  p^{(i)} & = & p_{\mu} \hat{e}_{(i)}^{\mu}, \quad i = r, \theta, \phi.
\end{eqnarray}
Thus,  we can obtain
\be
\bal
&\sin\alpha=\frac{p^{(\theta)}}{p^{(t)}}\\&=\left.\pm\frac{1}{ \hat{e}_{(t)}^t-L_E \hat{e}_{(t)}^\phi}
  \sqrt{\frac{\Delta_\theta K_E\sin^2\theta-(\chi-L_E)^2}{\Sigma \Delta_\theta\sin^2\theta}}\right|_{(r_O,\theta_O)},
\eal
\ee
\be
\bal
&\tan \beta  = \frac{p^{(\phi)}}{p^{(r)}}\\&=\left.\frac{L_E \sqrt{\Sigma \Delta_r}}
    {\sqrt{g_{\phi\phi}}\sqrt{\left[\left(\Sigma+a^2\sin^2\theta\right)^{2}-aL_E\right]^2-\Delta_rK_E}}\right|_{(r_O,\theta_O)},
\eal
\ee
where $\hat{e}_{(t)}^t$ and $\hat{e}_{(t)}^\phi$ are evaluated at the photon radius $r_{p}$. $r_{O}$ is the position of the observer and $\theta_{O}$ is the inclination angle between the observer’s line of sight and the rotation axis of the black hole. The  Cartesian celestial coordinate for the critical curve of the black hole shadow can be read off as
\be
  x \equiv - r_O \beta, \quad y \equiv r_O \alpha.
\ee

\begin{figure}[!htbp] 
   \centering 
   \includegraphics[width=2in]{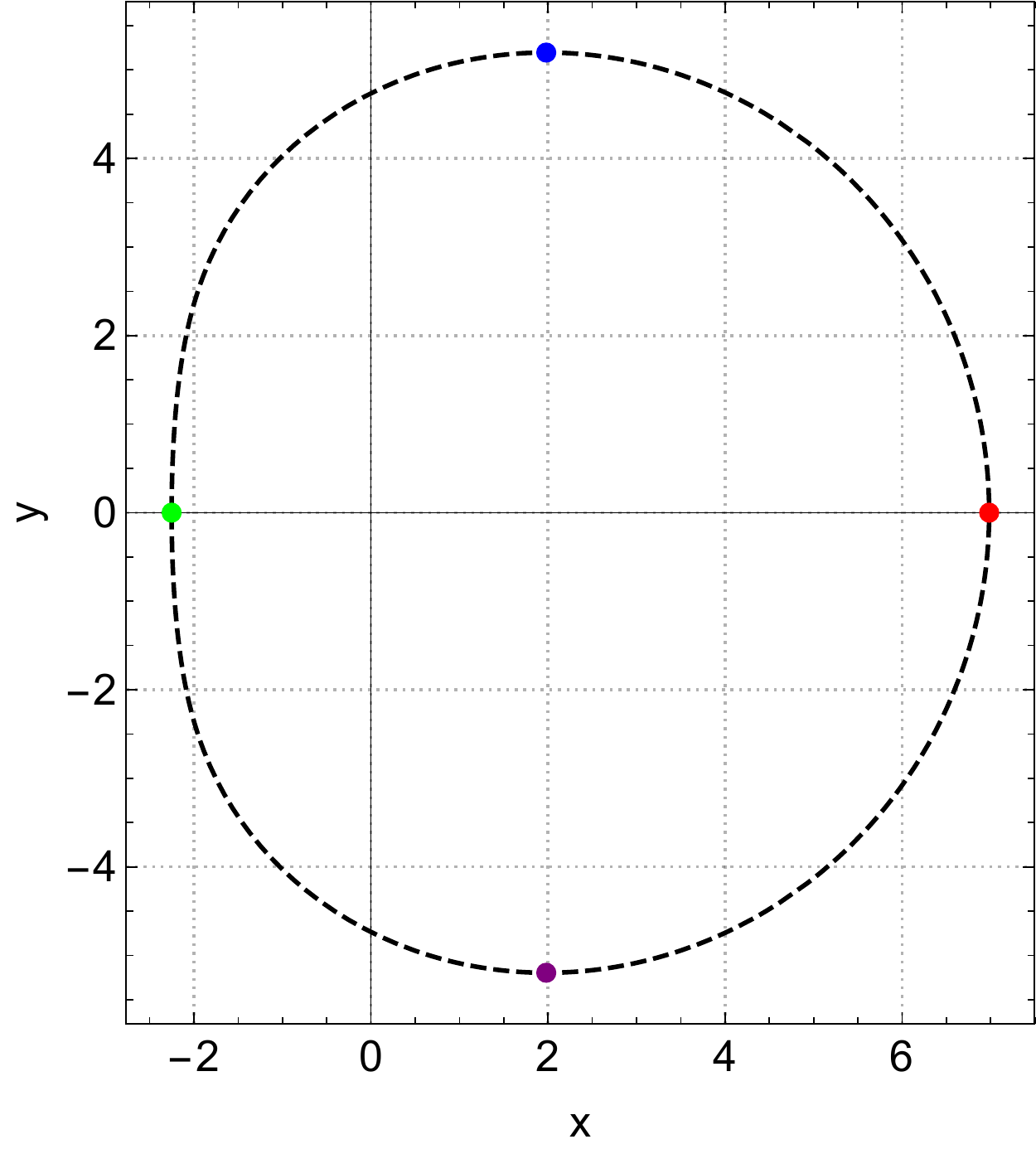}
   \caption{A schematic diagram of the black hole shadow. The coordinates of  the leftmost, rightmost, topmost, and lowest points are individually denoted by $(X_{l},~0),~(X_{r},~0),~(X_{t},~Y_{t}),$ and $(X_{t},~-Y_{t})$. }
   \label{figx1}
\end{figure}

According to the schematic diagram Fig. \ref{figx1}, the shadow size  $R_{s}$ of the black hole can be defined by \cite{Hioki:2009na}
\begin{equation}
R_{s}=\frac{(X_{t}-X_{r})^{2}+Y_{t}^{2}}{2(X_{r}-X_{t})}.
\end{equation}
We can also define the distortion $\delta_{s}$ of the black hole shadow as  \cite{Hioki:2009na}
\be
\delta_{s}=\frac{2 R_{s}-(X_{r}-X_{l})}{R_{s}}.
\ee

Generally, we have 
\begin{eqnarray}
  r_{h} & = &r_{h}(M,\,a\,,\Lambda), \\
  T & = & T(r_{h},\,a\,,\Lambda), \\
  R_{s} & = &R_{s}(M,\,a\,,\Lambda),\\
   \delta_{s} & = &\delta_{s}(M,\,a\,,\Lambda).
\end{eqnarray}
So the temperature of the Kerr black hole can be a function of not only the event horizon radius $r_{h}$, but also the size of the shadow $R_{s}$ or the distortion of the shadow $\delta_{s}$, {\it{i.e.}},
\begin{equation}
T(r_{h})=T(R_{s})=T(\delta_{s}).
\end{equation}
But even so, we can not know whether the size and the distortion of the shadow can be proper quantities reflecting the phase structures of the Kerr-(A)dS black hole.

\begin{figure}[!htbp] 
   \centering 
   ~\includegraphics[width=2.1in]{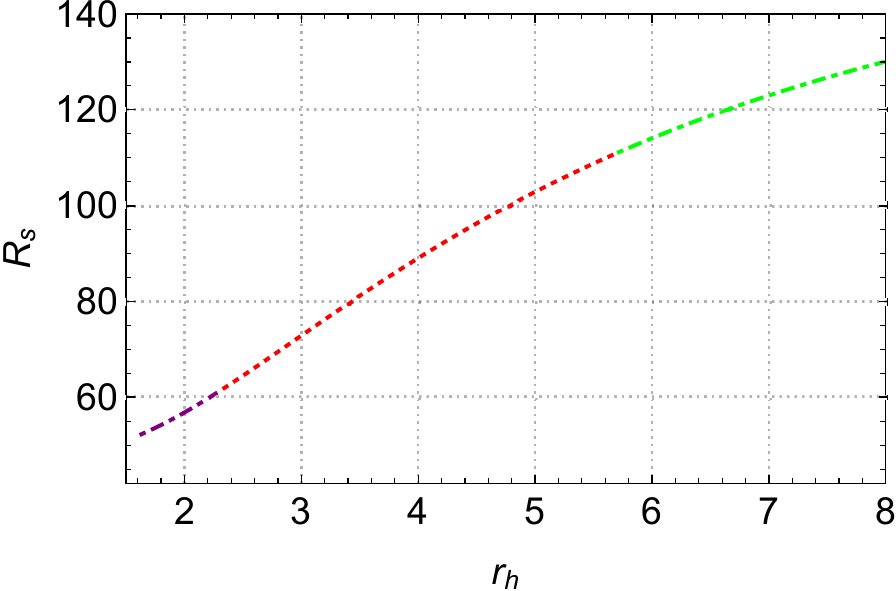}\\
   \includegraphics[width=2.25in]{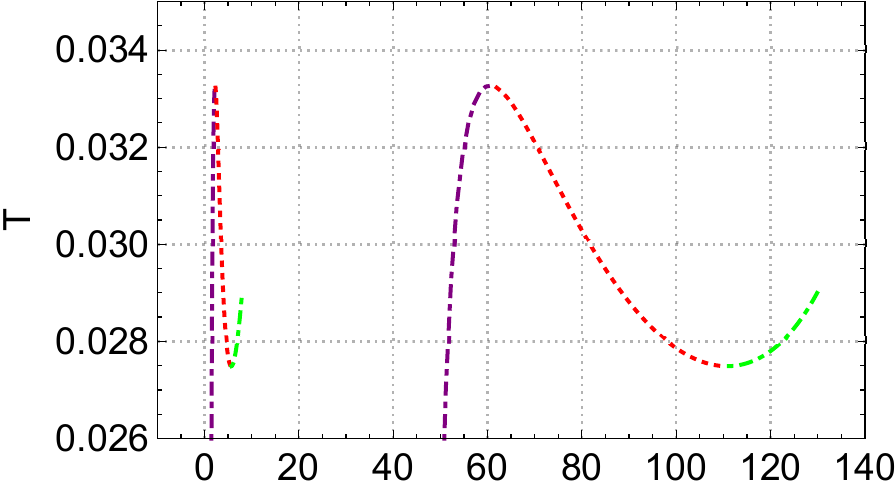}\,
   \caption{The upper diagram shows the relation between the shadow radius and the event horizon for the Kerr-AdS black hole. The bottom diagram shows the temperature in terms of the event horizon (left curve) and the shadow radius (right curve) for the black hole, respectively. We set $r_{O}=100,\,J=1,\,\theta_{O}=\pi/2,\, \Lambda=-0.03$. The purple, red and green curves in the top diagram correspond to the curves with the same colours in the bottom diagram.}
   \label{figx2}
\end{figure}

\begin{figure}[!htbp] 
   \centering 
   \includegraphics[width=2.1in]{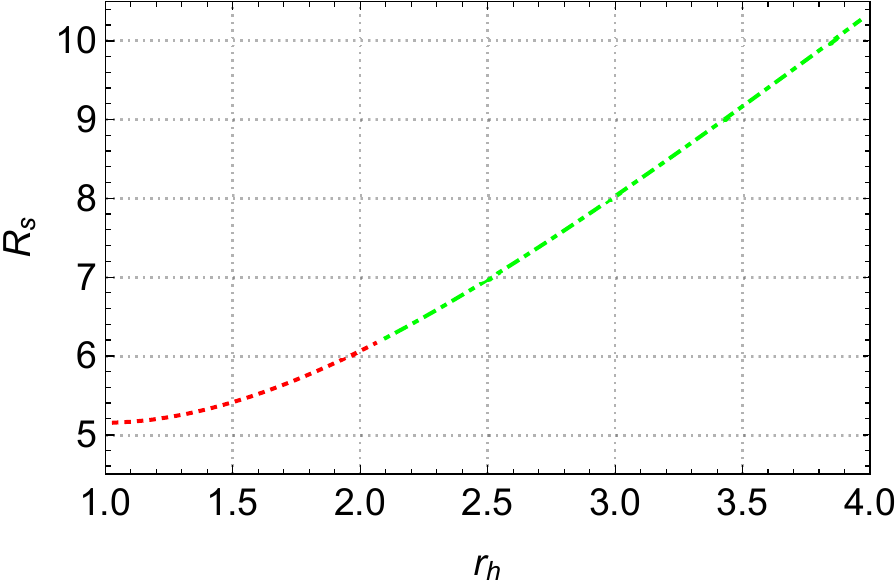}\\
   \includegraphics[width=2.17in]{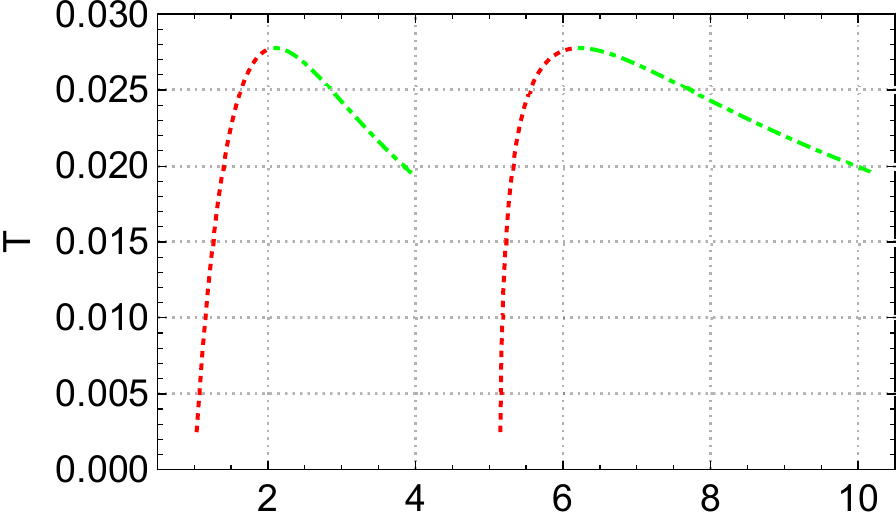}\,\,\,\,\,\,\,\,\,
   \caption{The upper diagram shows the relation between the shadow radius and the event horizon for the Kerr black hole. The bottom diagram shows the temperature in terms of the event horizon (left curve) and the shadow radius (right curve) for the black hole, respectively. We set $r_{O}=100,\,J=1,\,\theta_{O}=\pi/2,\, \Lambda=0$. The red and green curves in the top diagram correspond to the curves with the same colours in the bottom diagram.}
   \label{figx3}
\end{figure}

\begin{figure}[!htbp] 
   \centering 
    \includegraphics[width=2in]{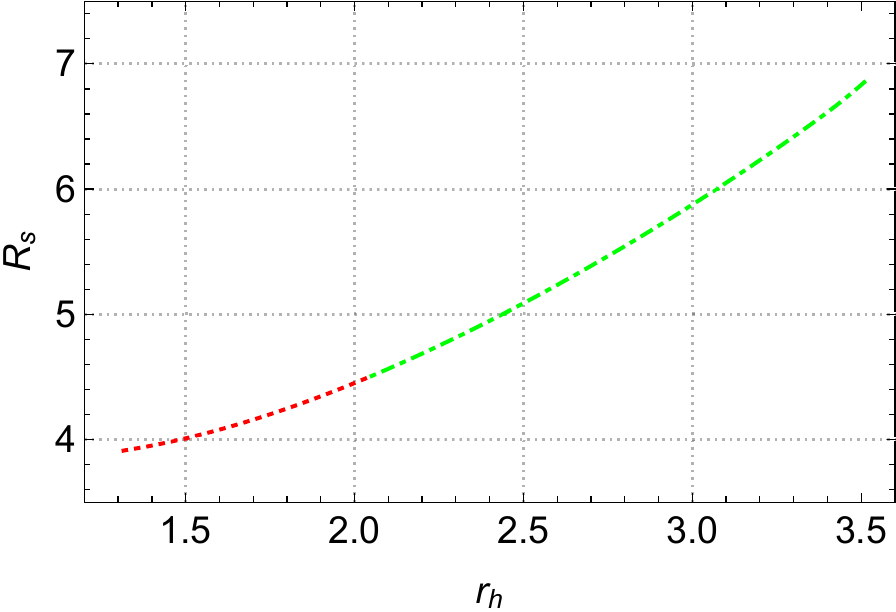}\\
 \includegraphics[width=2.19in]{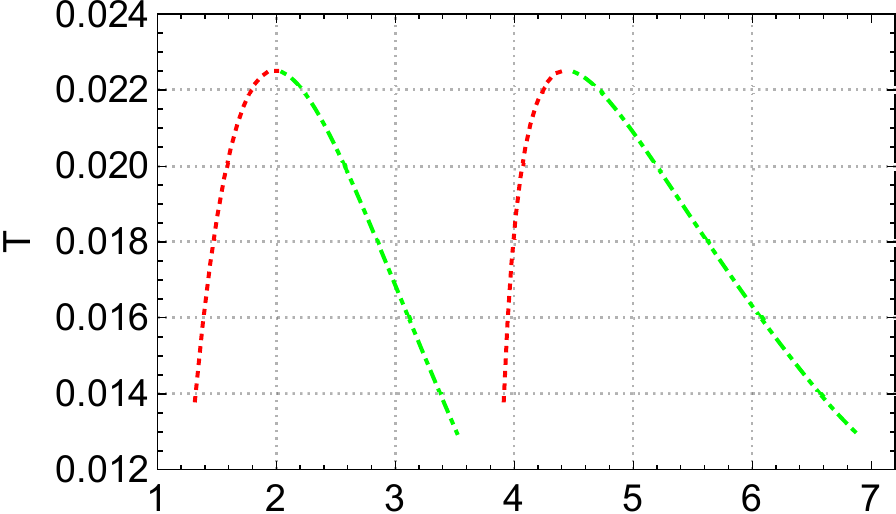}\,\,\,\,\,\,\,\,\,
   \caption{The upper diagram shows the relation between the shadow radius and the event horizon for the Kerr-dS black hole. The bottom diagram shows the temperature in terms of the event horizon (left curve) and the shadow radius (right curve) for the black hole, respectively.  We set $r_{O}=5,\,J=1,\,\theta_{O}=\pi/2, \,\Lambda=0.03$. The red and green curves in the top diagram correspond to  the curves with the same colours in the bottom diagram.}
   \label{figx4}
\end{figure}

\begin{figure}[!htbp] 
   \centering 
    \includegraphics[width=2.9in]{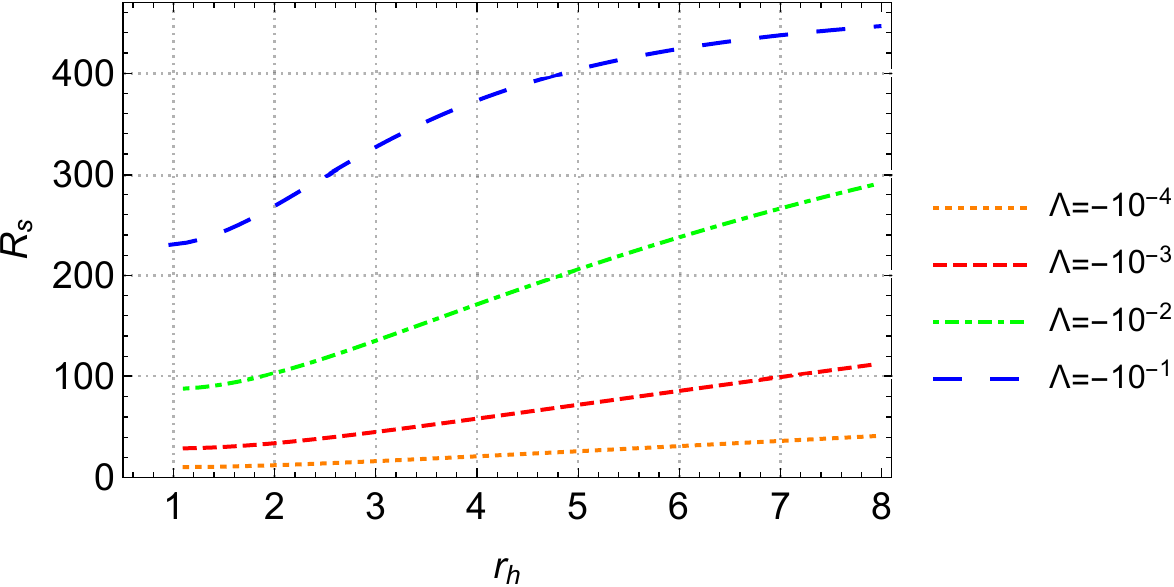}\\
 ~~~~~~~\includegraphics[width=3in]{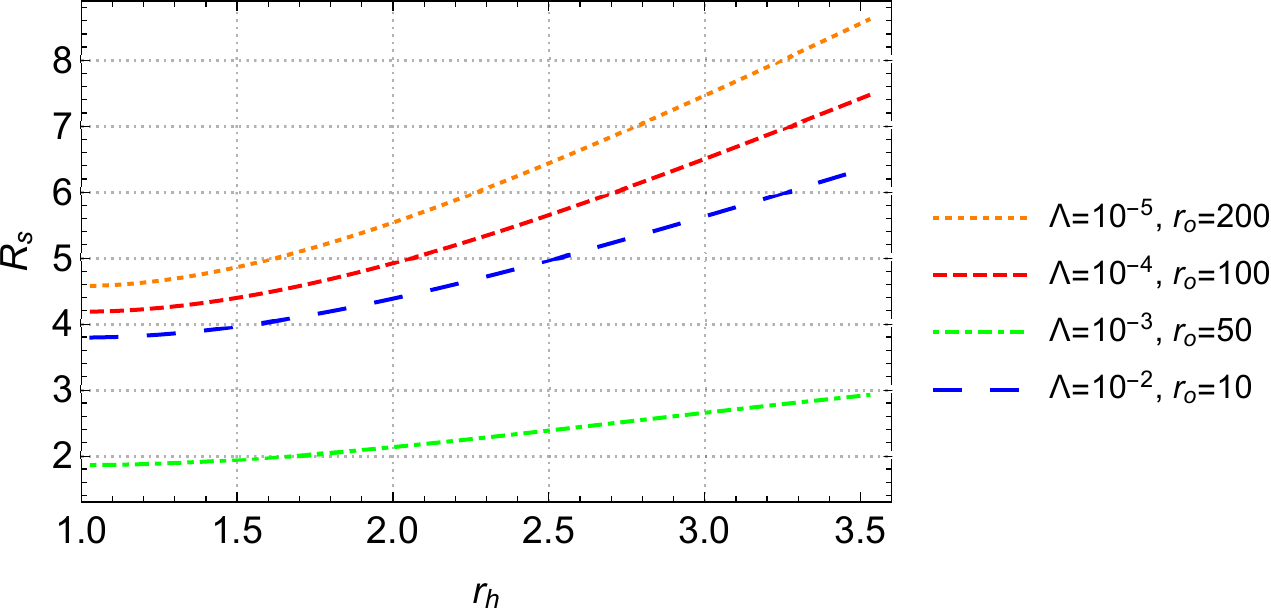}
   \caption{The variations of the shadow radii with respect to the event horizons for the Kerr-AdS black holes (top) and the Kerr-dS black holes (bottom). We have set $J=1,\,\theta_O=\pi/2,\,r_O=300$ for the top diagram and $J=1\,,\theta_O=\pi/2$ for the bottom diagram.}
   \label{figx6}
\end{figure}

\begin{figure*}[!htbp] 
   \centering 
    \includegraphics[width=2.19in]{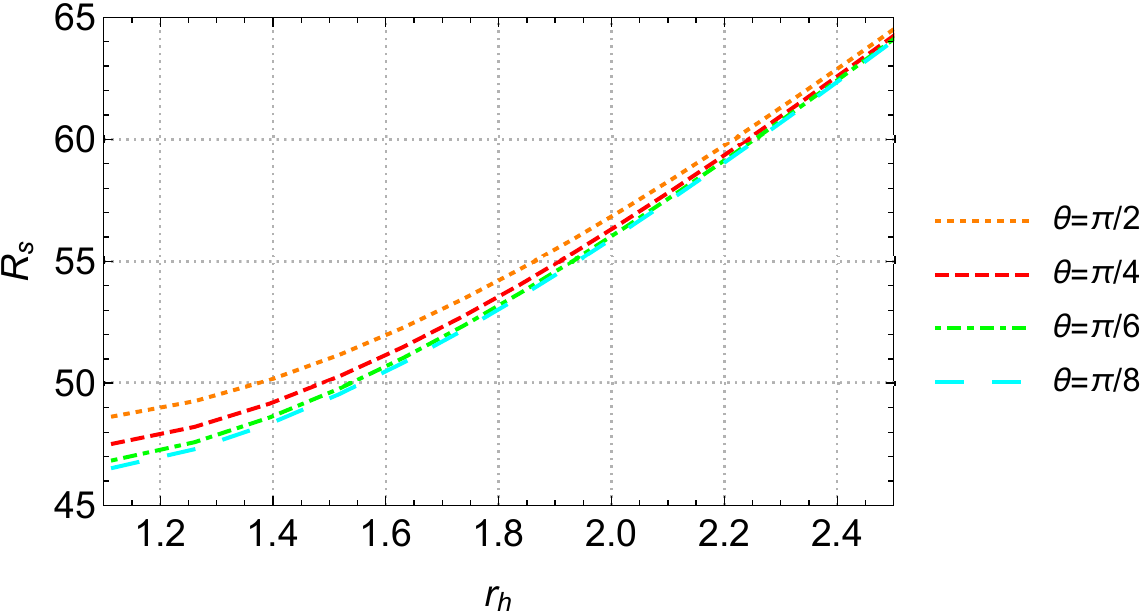}
 \includegraphics[width=2.25in]{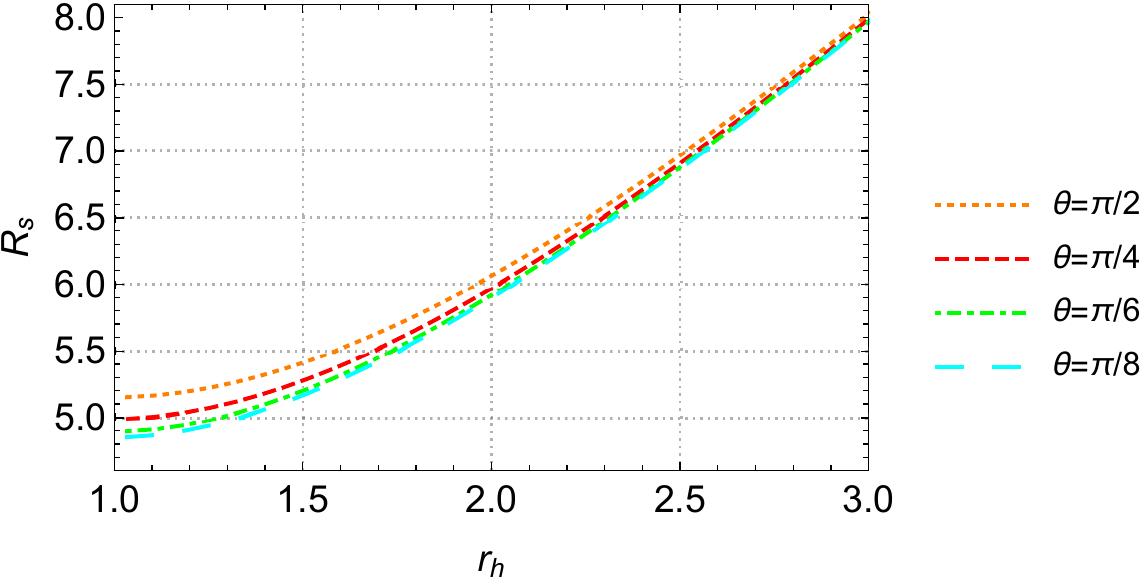}
  \includegraphics[width=2.2in]{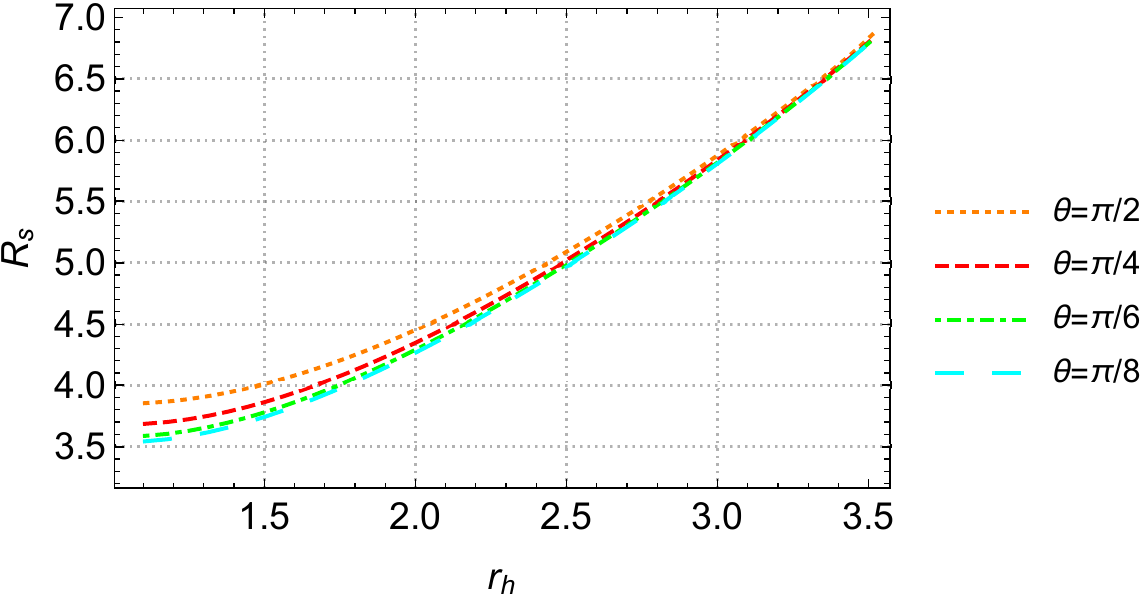}
   \caption{The variations of the shadow radii with respect to the event horizons for the Kerr-AdS black holes (left, $J=1,\, r_O=300,\,\Lambda=-0.03$), the Kerr black holes (middle, $J=1,\, r_O=100,\,\Lambda=0$) and the Kerr-dS black holes (right, $J=1, \,r_O=5,\,\Lambda=0.03$). }
   \label{figx7}
\end{figure*}

\begin{figure*}[!htbp] 
   \centering 
    \includegraphics[width=2.19in]{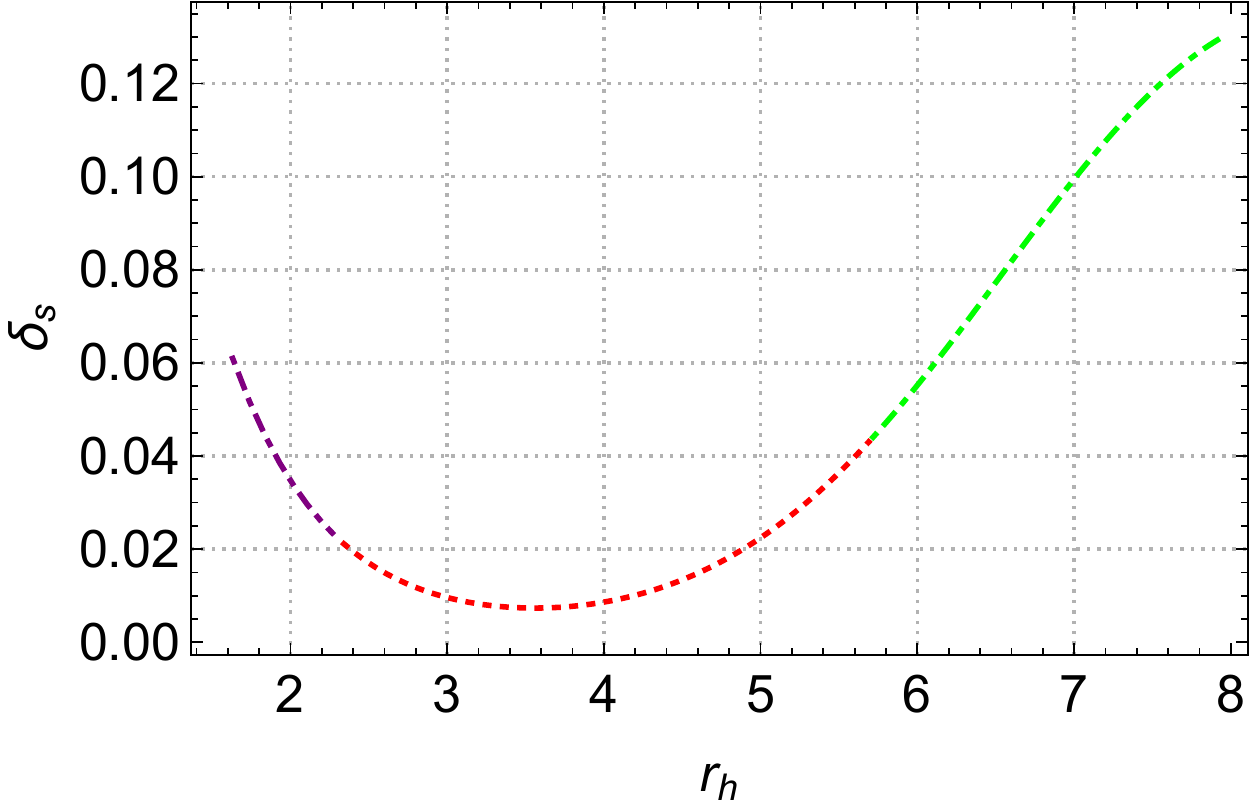}
 \includegraphics[width=2.25in]{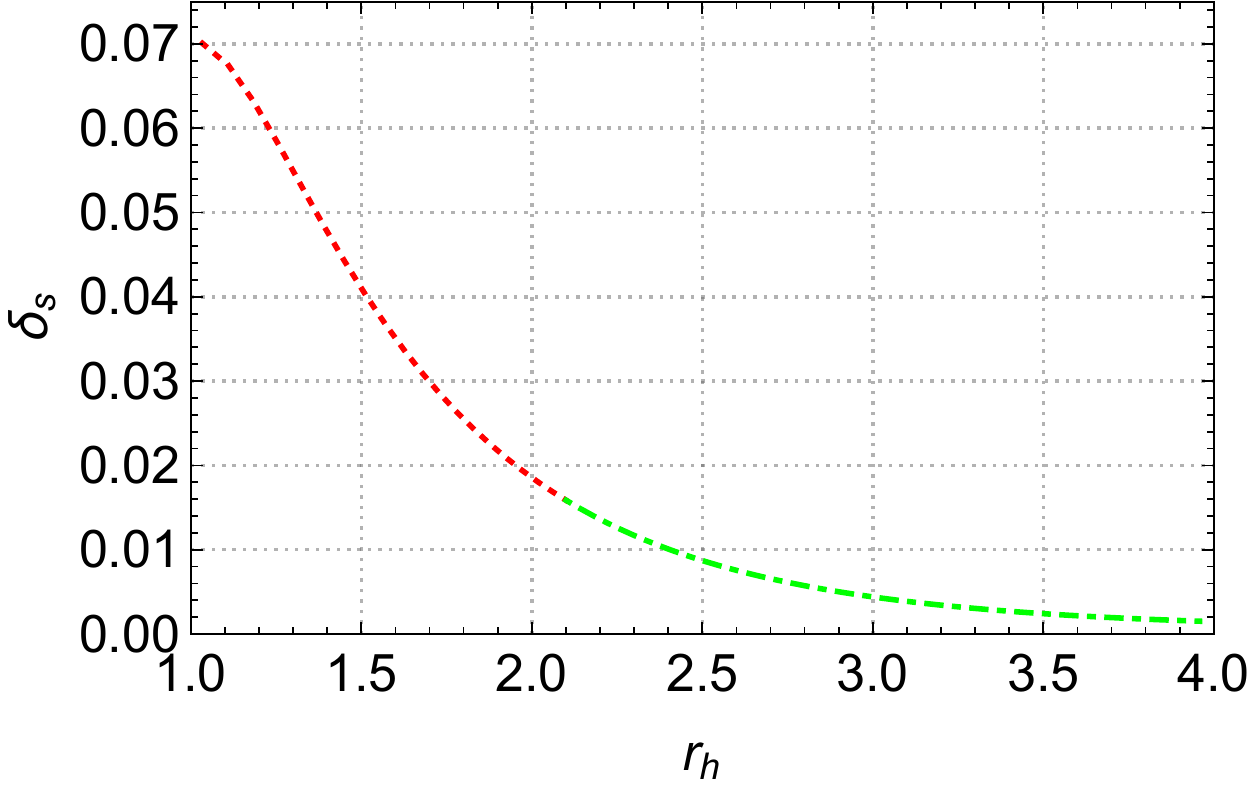}
  \includegraphics[width=2.2in]{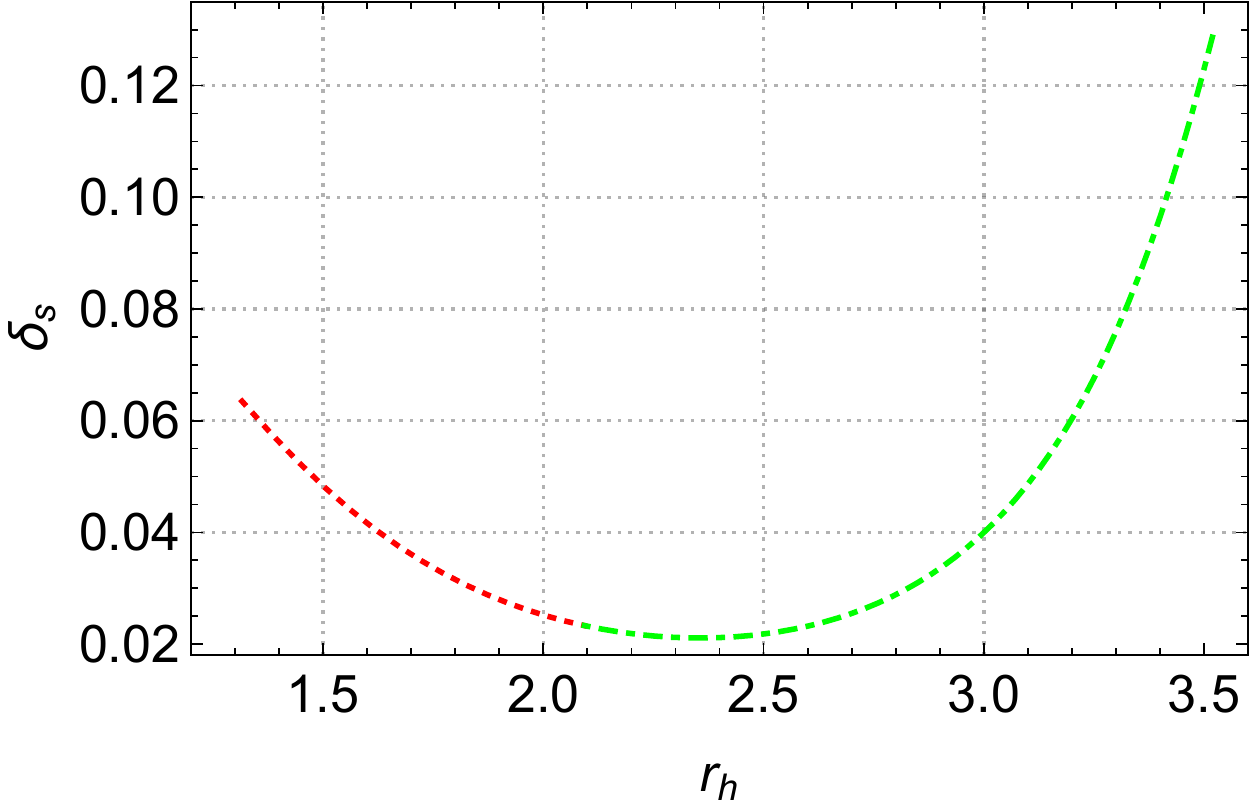}
   \caption{The variations of the distortions for the Kerr-AdS, Kerr and Kerr-dS black holes with respect to the event horizons of the black holes. The parameters are set to be $J=1,\,\Lambda=-0.03,\,\theta_O=\pi/4$ for the left panel, $J=1,\,\Lambda=0,\,\theta_O=\pi/6$ for the middle panel and $J=1,\,\Lambda=0.03,\,\theta_O=\pi/6$ for the right panel.}
   \label{figx5}
\end{figure*}

We have shown the relations between the event horizon $r_{h}$ and the size of the shadow $R_{s}$ for the Kerr-AdS black hole, the Kerr black hole and the Kerr-dS black hole in upper diagrams of Figs. \ref{figx2}, \ref{figx3} and \ref{figx4}, respectively. We can see that
\begin{equation}\label{monors}
\frac{dR_{s}}{dr_{h}}>0
\end{equation}
for all cases.
As a result, we can know that
\begin{equation}\label{kerreone}
\frac{dT}{dr_{h}}>0,~~\frac{dT}{dr_{h}}=0,~~\frac{dT}{dr_{h}}<0
\end{equation}
correspond to
\begin{equation}\label{kerretwo}
\frac{dT}{dR_{s}}>0,~~\frac{dT}{dR_{s}}=0,~~\frac{dT}{dR_{s}}<0.
\end{equation}
We can observe this clearly in the bottom diagrams of Figs. \ref{figx2}, \ref{figx3} and \ref{figx4} for the Kerr black hole with a negative cosmological constant, zero cosmological constant and a positive cosmological constant. To warrant the universality of the property, we furthermore show the relations between the shadow radii and the event horizons for the Kerr -(A)dS black holes with different cosmological constants in Fig. \ref{figx6} and different inclination angles in Fig. \ref{figx7}. It is clear that all these cases do comply with (\ref{monors}).  Therefore, similar to the generic spherically symmetric black hole case, the thermodynamic phase structure of the axially symmetric Kerr-(A)dS black hole can be reflected by the size of its shadow. 

We may then wonder whether the distortion of the shadow shares this property. We further show the relations between the event horizon $r_{h}$ and the distortion of the shadow $\delta_{s}$ in Fig. \ref{figx5}. Unfortunately, we can not always have $d\delta_{s}/dr_{h}>0$. Consequently, the thermodynamic phase structures of the axially symmetric Kerr black hole (with a cosmological constant) can not be reflected by the distortion of its shadow.

Here we would like to briefly talk about the selections of the parameters. For all the cases, we should have $r_O>r_p$. For the observer in the Kerr-dS spacetime, we need $r_O<r_C$ with $r_C$ the cosmological horizon. According to the first law of the black hole thermodynamics, we obtained the curves in Figs. (\ref{figx2})-(\ref{figx5})  by setting the angular momentum $J$ of the Kerr-(A)dS black hole as a constant and varying the entropy. For instance, in the Fig. \ref{figx3} and the middle panel of Fig. \ref{figx5}, with the constant angular momentum $J=1$, the entropy $S$ of the Kerr black hole varies from $6.5$ to $50$ monotonically, the corresponding event horizon $r_h$ varies from $1.034$ to $3.958$ and it coincides with the formula for the event horizon radius $r_h=m+\sqrt{m^2-a^2}$ for the Kerr black hole.

\section{Conclusions and Discussions}\label{remarks}
Previously, the black hole shadow and the black hole thermodynamics were studied separately; of particular interest is the possible relation between them, which is the topic in this paper. Firstly, we found that the radius of the shadow can be a well-defined characteristic quantity to reflect the phase structure of the spherically symmetric black hole.  The shadow radius of the spherically symmetric black hole is closely related to the photon radius around the black hole by (\ref{shara}). We observe the connection (\ref{dre3}) between the shadow radius and the event horizon of the spherically symmetric black hole using the photon radius.

Secondly, we showed that the size of the shadow can reflect the phase structure of the axially symmetric black hole. We did this by choosing the Kerr black hole (with a cosmological constant)  as a toy model. We have tested cases of $\Lambda<0,\,\Lambda=0$ and $\Lambda>0$ and explored the relations of the shadow size and the black hole horizon by numerical calculations, obtaining $dR_{s}/dr_{h}$ for all cases. We have also tried to use a direct definition \cite{Feng:2019zzn} of the black hole shadow, $R=(X_{r}-X_{l})/2$, to reflect the size of the black hole, and it works well too. However, we found that the distortion of the shadow can not give proper information on the phase structures for the Kerr black hole with a cosmological constant.

On the one hand, in fact, both the event horizon of the black hole and the size of the black hole shadow directly depend on the mass of the black hole. According to the thermodynamic first law of the black hole, we can know that the variations of the black hole phases are in fact correspond to the variations of the black hole mass. So it is not difficult to understand that both the event horizon and the shadow size can reflect the phase structures of the black hole. One the other hand, the shadow distortion's failure of being an appropriate quantity reflecting the phase structure of the axially symmetric black hole, we think, is due to irregularity and complexity of the shadow shape contour.

\section*{Acknowledgements}
The authors thank Haopeng Yan for inspired discussions and comments. MG is supported by NSFC Grant No. 11947210.
And he is also funded by China Postdoctoral Science Foundation Grant No. 2019M660278 and 2020T130020. M. Zhang is supported by the Initial Research Foundation of Jiangxi Normal University with Grant No. 12020023.

\appendix
\section{Some thermodynamic quantities of Kerr-(A)dS black hole}\label{ap}
The Smarr relation for the Kerr-AdS black hole is \cite{Caldarelli:1999xj}
\begin{equation}
M=\sqrt{\frac{J^2}{l^2}+\frac{\pi  J^2}{S}+\frac{S^3}{4 \pi ^3 l^4}+\frac{S^2}{2 \pi ^2 l^2}+\frac{S}{4 \pi }},
\end{equation}
where $l$ is the AdS radius relating to the cosmological constant by $\Lambda =-3/l^{2}$, $M$ is the mass of the black hole and it is related to the mass parameter $m$ by \cite{Gibbons:2004ai,Cvetic:2010jb,Gunasekaran:2012dq}
\begin{equation}
M=\frac{m}{\left(1-\frac{a^{2}}{l^{2}}\right)^{2}}.
\end{equation}
$S$ is the Bekenstein-Hawking entropy of the black hole and in terms of the event horizon we have
\begin{equation}
S=\frac{\pi \left(r_{h}^{2}+a^{2}\right)}{1-\frac{a^{2}}{l^{2}}}.
\end{equation}
The Smarr relation for the Kerr-dS black hole on the event horizon is
\begin{equation}
M=\sqrt{-\frac{J^2}{l^2}+\frac{\pi  J^2}{S}-\frac{S^3}{4 \pi ^3 l^4}-\frac{S^2}{2 \pi ^2 l^2}+\frac{S}{4 \pi }},
\end{equation}
where $l$ is the dS radius and $\Lambda =3/l^{2}$. The mass and the entropy of the black hole are \cite{Dolan:2013ft}
\begin{align}
M=&\frac{m}{\left(1+\frac{a^{2}}{l^{2}}\right)^{2}},\\S=&\frac{\pi \left(r_{h}^{2}+a^{2}\right)}{1+\frac{a^{2}}{l^{2}}}.
\end{align}
We do not consider the thermodynamic quantities and phase structures on the cosmological horizon of the black hole.


\begin{thebibliography}{52}%
\makeatletter
\providecommand \@ifxundefined [1]{%
 \@ifx{#1\undefined}
}%
\providecommand \@ifnum [1]{%
 \ifnum #1\expandafter \@firstoftwo
 \else \expandafter \@secondoftwo
 \fi
}%
\providecommand \@ifx [1]{%
 \ifx #1\expandafter \@firstoftwo
 \else \expandafter \@secondoftwo
 \fi
}%
\providecommand \natexlab [1]{#1}%
\providecommand \enquote  [1]{``#1''}%
\providecommand \bibnamefont  [1]{#1}%
\providecommand \bibfnamefont [1]{#1}%
\providecommand \citenamefont [1]{#1}%
\providecommand \href@noop [0]{\@secondoftwo}%
\providecommand \href [0]{\begingroup \@sanitize@url \@href}%
\providecommand \@href[1]{\@@startlink{#1}\@@href}%
\providecommand \@@href[1]{\endgroup#1\@@endlink}%
\providecommand \@sanitize@url [0]{\catcode `\\12\catcode `\$12\catcode
  `\&12\catcode `\#12\catcode `\^12\catcode `\_12\catcode `\%12\relax}%
\providecommand \@@startlink[1]{}%
\providecommand \@@endlink[0]{}%
\providecommand \url  [0]{\begingroup\@sanitize@url \@url }%
\providecommand \@url [1]{\endgroup\@href {#1}{\urlprefix }}%
\providecommand \urlprefix  [0]{URL }%
\providecommand \Eprint [0]{\href }%
\providecommand \doibase [0]{http://dx.doi.org/}%
\providecommand \selectlanguage [0]{\@gobble}%
\providecommand \bibinfo  [0]{\@secondoftwo}%
\providecommand \bibfield  [0]{\@secondoftwo}%
\providecommand \translation [1]{[#1]}%
\providecommand \BibitemOpen [0]{}%
\providecommand \bibitemStop [0]{}%
\providecommand \bibitemNoStop [0]{.\EOS\space}%
\providecommand \EOS [0]{\spacefactor3000\relax}%
\providecommand \BibitemShut  [1]{\csname bibitem#1\endcsname}%
\let\auto@bib@innerbib\@empty
\bibitem [{\citenamefont {Akiyama}\ \emph
  {et~al.}(2019{\natexlab{a}})\citenamefont {Akiyama} \emph
  {et~al.}}]{Akiyama:2019cqa}%
  \BibitemOpen
  \bibfield  {author} {\bibinfo {author} {\bibfnamefont {K.}~\bibnamefont
  {Akiyama}} \emph {et~al.} (\bibinfo {collaboration} {Event Horizon
  Telescope}),\ }\href {\doibase 10.3847/2041-8213/ab0ec7} {\bibfield
  {journal} {\bibinfo  {journal} {Astrophys. J.}\ }\textbf {\bibinfo {volume}
  {875}},\ \bibinfo {pages} {L1} (\bibinfo {year}
  {2019}{\natexlab{a}})}\BibitemShut {NoStop}%
\bibitem [{\citenamefont {Akiyama}\ \emph
  {et~al.}(2019{\natexlab{b}})\citenamefont {Akiyama} \emph
  {et~al.}}]{Akiyama:2019brx}%
  \BibitemOpen
  \bibfield  {author} {\bibinfo {author} {\bibfnamefont {K.}~\bibnamefont
  {Akiyama}} \emph {et~al.} (\bibinfo {collaboration} {Event Horizon
  Telescope}),\ }\href {\doibase 10.3847/2041-8213/ab0c96} {\bibfield
  {journal} {\bibinfo  {journal} {Astrophys. J.}\ }\textbf {\bibinfo {volume}
  {875}},\ \bibinfo {pages} {L2} (\bibinfo {year}
  {2019}{\natexlab{b}})}\BibitemShut {NoStop}%
\bibitem [{\citenamefont {Akiyama}\ \emph
  {et~al.}(2019{\natexlab{c}})\citenamefont {Akiyama} \emph
  {et~al.}}]{Akiyama:2019sww}%
  \BibitemOpen
  \bibfield  {author} {\bibinfo {author} {\bibfnamefont {K.}~\bibnamefont
  {Akiyama}} \emph {et~al.} (\bibinfo {collaboration} {Event Horizon
  Telescope}),\ }\href {\doibase 10.3847/2041-8213/ab0c57} {\bibfield
  {journal} {\bibinfo  {journal} {Astrophys. J.}\ }\textbf {\bibinfo {volume}
  {875}},\ \bibinfo {pages} {L3} (\bibinfo {year}
  {2019}{\natexlab{c}})}\BibitemShut {NoStop}%
\bibitem [{\citenamefont {Akiyama}\ \emph
  {et~al.}(2019{\natexlab{d}})\citenamefont {Akiyama} \emph
  {et~al.}}]{Akiyama:2019bqs}%
  \BibitemOpen
  \bibfield  {author} {\bibinfo {author} {\bibfnamefont {K.}~\bibnamefont
  {Akiyama}} \emph {et~al.} (\bibinfo {collaboration} {Event Horizon
  Telescope}),\ }\href {\doibase 10.3847/2041-8213/ab0e85} {\bibfield
  {journal} {\bibinfo  {journal} {Astrophys. J.}\ }\textbf {\bibinfo {volume}
  {875}},\ \bibinfo {pages} {L4} (\bibinfo {year}
  {2019}{\natexlab{d}})}\BibitemShut {NoStop}%
\bibitem [{\citenamefont {Akiyama}\ \emph
  {et~al.}(2019{\natexlab{e}})\citenamefont {Akiyama} \emph
  {et~al.}}]{Akiyama:2019fyp}%
  \BibitemOpen
  \bibfield  {author} {\bibinfo {author} {\bibfnamefont {K.}~\bibnamefont
  {Akiyama}} \emph {et~al.} (\bibinfo {collaboration} {Event Horizon
  Telescope}),\ }\href {\doibase 10.3847/2041-8213/ab0f43} {\bibfield
  {journal} {\bibinfo  {journal} {Astrophys. J.}\ }\textbf {\bibinfo {volume}
  {875}},\ \bibinfo {pages} {L5} (\bibinfo {year}
  {2019}{\natexlab{e}})}\BibitemShut {NoStop}%
\bibitem [{\citenamefont {Akiyama}\ \emph
  {et~al.}(2019{\natexlab{f}})\citenamefont {Akiyama} \emph
  {et~al.}}]{Akiyama:2019eap}%
  \BibitemOpen
  \bibfield  {author} {\bibinfo {author} {\bibfnamefont {K.}~\bibnamefont
  {Akiyama}} \emph {et~al.} (\bibinfo {collaboration} {Event Horizon
  Telescope}),\ }\href {\doibase 10.3847/2041-8213/ab1141} {\bibfield
  {journal} {\bibinfo  {journal} {Astrophys. J.}\ }\textbf {\bibinfo {volume}
  {875}},\ \bibinfo {pages} {L6} (\bibinfo {year}
  {2019}{\natexlab{f}})}\BibitemShut {NoStop}%
\bibitem [{\citenamefont {Synge}(1966)}]{synge1966escape}%
  \BibitemOpen
  \bibfield  {author} {\bibinfo {author} {\bibfnamefont {J.}~\bibnamefont
  {Synge}},\ }\href@noop {} {\bibfield  {journal} {\bibinfo  {journal} {Monthly
  Notices of the Royal Astronomical Society}\ }\textbf {\bibinfo {volume}
  {131}},\ \bibinfo {pages} {463} (\bibinfo {year} {1966})}\BibitemShut
  {NoStop}%
\bibitem [{\citenamefont {Luminet}(1979)}]{luminet1979image}%
  \BibitemOpen
  \bibfield  {author} {\bibinfo {author} {\bibfnamefont {J.-P.}\ \bibnamefont
  {Luminet}},\ }\href@noop {} {\bibfield  {journal} {\bibinfo  {journal}
  {Astronomy and Astrophysics}\ }\textbf {\bibinfo {volume} {75}},\ \bibinfo
  {pages} {228} (\bibinfo {year} {1979})}\BibitemShut {NoStop}%
\bibitem [{\citenamefont {Hawking}\ \emph {et~al.}(1973)\citenamefont
  {Hawking}, \citenamefont {Carter}, \citenamefont {Bardeen}, \citenamefont
  {Gursky}, \citenamefont {Thorne}, \citenamefont {Ruffini}, \citenamefont
  {Novikov} \emph {et~al.}}]{hawking1973black}%
  \BibitemOpen
  \bibfield  {author} {\bibinfo {author} {\bibfnamefont {S.}~\bibnamefont
  {Hawking}}, \bibinfo {author} {\bibfnamefont {B.}~\bibnamefont {Carter}},
  \bibinfo {author} {\bibfnamefont {J.~M.}\ \bibnamefont {Bardeen}}, \bibinfo
  {author} {\bibfnamefont {H.}~\bibnamefont {Gursky}}, \bibinfo {author}
  {\bibfnamefont {K.~S.}\ \bibnamefont {Thorne}}, \bibinfo {author}
  {\bibfnamefont {R.}~\bibnamefont {Ruffini}}, \bibinfo {author} {\bibfnamefont
  {I.~D.}\ \bibnamefont {Novikov}},  \emph {et~al.},\ }\href@noop {} {\emph
  {\bibinfo {title} {Black holes}}},\ Vol.~\bibinfo {volume} {23}\ (\bibinfo
  {publisher} {CRC Press},\ \bibinfo {year} {1973})\BibitemShut {NoStop}%
\bibitem [{\citenamefont {De~Vries}(2000)}]{de2000apparent}%
  \BibitemOpen
  \bibfield  {author} {\bibinfo {author} {\bibfnamefont {A.}~\bibnamefont
  {De~Vries}},\ }\href@noop {} {\bibfield  {journal} {\bibinfo  {journal}
  {Classical and Quantum Gravity}\ }\textbf {\bibinfo {volume} {17}},\ \bibinfo
  {pages} {123} (\bibinfo {year} {2000})}\BibitemShut {NoStop}%
\bibitem [{\citenamefont {Hioki}\ and\ \citenamefont
  {Miyamoto}(2008)}]{Hioki:2008zw}%
  \BibitemOpen
  \bibfield  {author} {\bibinfo {author} {\bibfnamefont {K.}~\bibnamefont
  {Hioki}}\ and\ \bibinfo {author} {\bibfnamefont {U.}~\bibnamefont
  {Miyamoto}},\ }\href {\doibase 10.1103/PhysRevD.78.044007} {\bibfield
  {journal} {\bibinfo  {journal} {Phys. Rev.}\ }\textbf {\bibinfo {volume}
  {D78}},\ \bibinfo {pages} {044007} (\bibinfo {year} {2008})},\ \Eprint
  {http://arxiv.org/abs/0805.3146} {arXiv:0805.3146 [gr-qc]} \BibitemShut
  {NoStop}%
\bibitem [{\citenamefont {Grenzebach}\ \emph {et~al.}(2014)\citenamefont
  {Grenzebach}, \citenamefont {Perlick},\ and\ \citenamefont
  {Lämmerzahl}}]{Grenzebach:2014fha}%
  \BibitemOpen
  \bibfield  {author} {\bibinfo {author} {\bibfnamefont {A.}~\bibnamefont
  {Grenzebach}}, \bibinfo {author} {\bibfnamefont {V.}~\bibnamefont {Perlick}},
  \ and\ \bibinfo {author} {\bibfnamefont {C.}~\bibnamefont {Lämmerzahl}},\
  }\href {\doibase 10.1103/PhysRevD.89.124004} {\bibfield  {journal} {\bibinfo
  {journal} {Phys. Rev.}\ }\textbf {\bibinfo {volume} {D89}},\ \bibinfo {pages}
  {124004} (\bibinfo {year} {2014})},\ \Eprint {http://arxiv.org/abs/1403.5234}
  {arXiv:1403.5234 [gr-qc]} \BibitemShut {NoStop}%
\bibitem [{\citenamefont {Wang}\ \emph {et~al.}(2018)\citenamefont {Wang},
  \citenamefont {Chen},\ and\ \citenamefont {Jing}}]{Wang:2017qhh}%
  \BibitemOpen
  \bibfield  {author} {\bibinfo {author} {\bibfnamefont {M.}~\bibnamefont
  {Wang}}, \bibinfo {author} {\bibfnamefont {S.}~\bibnamefont {Chen}}, \ and\
  \bibinfo {author} {\bibfnamefont {J.}~\bibnamefont {Jing}},\ }\href {\doibase
  10.1103/PhysRevD.97.064029} {\bibfield  {journal} {\bibinfo  {journal} {Phys.
  Rev.}\ }\textbf {\bibinfo {volume} {D97}},\ \bibinfo {pages} {064029}
  (\bibinfo {year} {2018})},\ \Eprint {http://arxiv.org/abs/1710.07172}
  {arXiv:1710.07172 [gr-qc]} \BibitemShut {NoStop}%
\bibitem [{\citenamefont {Guo}\ \emph {et~al.}(2018)\citenamefont {Guo},
  \citenamefont {Obers},\ and\ \citenamefont {Yan}}]{Guo:2018kis}%
  \BibitemOpen
  \bibfield  {author} {\bibinfo {author} {\bibfnamefont {M.}~\bibnamefont
  {Guo}}, \bibinfo {author} {\bibfnamefont {N.~A.}\ \bibnamefont {Obers}}, \
  and\ \bibinfo {author} {\bibfnamefont {H.}~\bibnamefont {Yan}},\ }\href
  {\doibase 10.1103/PhysRevD.98.084063} {\bibfield  {journal} {\bibinfo
  {journal} {Phys. Rev.}\ }\textbf {\bibinfo {volume} {D98}},\ \bibinfo {pages}
  {084063} (\bibinfo {year} {2018})},\ \Eprint
  {http://arxiv.org/abs/1806.05249} {arXiv:1806.05249 [gr-qc]} \BibitemShut
  {NoStop}%
\bibitem [{\citenamefont {Yan}(2019)}]{Yan:2019etp}%
  \BibitemOpen
  \bibfield  {author} {\bibinfo {author} {\bibfnamefont {H.}~\bibnamefont
  {Yan}},\ }\href {\doibase 10.1103/PhysRevD.99.084050} {\bibfield  {journal}
  {\bibinfo  {journal} {Phys. Rev.}\ }\textbf {\bibinfo {volume} {D99}},\
  \bibinfo {pages} {084050} (\bibinfo {year} {2019})},\ \Eprint
  {http://arxiv.org/abs/1903.04382} {arXiv:1903.04382 [gr-qc]} \BibitemShut
  {NoStop}%
\bibitem [{\citenamefont {Hennigar}\ \emph {et~al.}(2018)\citenamefont
  {Hennigar}, \citenamefont {Poshteh},\ and\ \citenamefont
  {Mann}}]{Hennigar:2018hza}%
  \BibitemOpen
  \bibfield  {author} {\bibinfo {author} {\bibfnamefont {R.~A.}\ \bibnamefont
  {Hennigar}}, \bibinfo {author} {\bibfnamefont {M.~B.~J.}\ \bibnamefont
  {Poshteh}}, \ and\ \bibinfo {author} {\bibfnamefont {R.~B.}\ \bibnamefont
  {Mann}},\ }\href {\doibase 10.1103/PhysRevD.97.064041} {\bibfield  {journal}
  {\bibinfo  {journal} {Phys. Rev.}\ }\textbf {\bibinfo {volume} {D97}},\
  \bibinfo {pages} {064041} (\bibinfo {year} {2018})},\ \Eprint
  {http://arxiv.org/abs/1801.03223} {arXiv:1801.03223 [gr-qc]} \BibitemShut
  {NoStop}%
\bibitem [{\citenamefont {Konoplya}(2019)}]{Konoplya:2019sns}%
  \BibitemOpen
  \bibfield  {author} {\bibinfo {author} {\bibfnamefont {R.~A.}\ \bibnamefont
  {Konoplya}},\ }\href {\doibase 10.1016/j.physletb.2019.05.043} {\bibfield
  {journal} {\bibinfo  {journal} {Phys. Lett.}\ }\textbf {\bibinfo {volume}
  {B795}},\ \bibinfo {pages} {1} (\bibinfo {year} {2019})},\ \Eprint
  {http://arxiv.org/abs/1905.00064} {arXiv:1905.00064 [gr-qc]} \BibitemShut
  {NoStop}%
\bibitem [{\citenamefont {Bambi}\ and\ \citenamefont
  {Freese}(2009)}]{Bambi:2008jg}%
  \BibitemOpen
  \bibfield  {author} {\bibinfo {author} {\bibfnamefont {C.}~\bibnamefont
  {Bambi}}\ and\ \bibinfo {author} {\bibfnamefont {K.}~\bibnamefont {Freese}},\
  }\href {\doibase 10.1103/PhysRevD.79.043002} {\bibfield  {journal} {\bibinfo
  {journal} {Phys. Rev.}\ }\textbf {\bibinfo {volume} {D79}},\ \bibinfo {pages}
  {043002} (\bibinfo {year} {2009})},\ \Eprint {http://arxiv.org/abs/0812.1328}
  {arXiv:0812.1328 [astro-ph]} \BibitemShut {NoStop}%
\bibitem [{\citenamefont {Amir}\ \emph {et~al.}(2018)\citenamefont {Amir},
  \citenamefont {Singh},\ and\ \citenamefont {Ghosh}}]{Amir:2017slq}%
  \BibitemOpen
  \bibfield  {author} {\bibinfo {author} {\bibfnamefont {M.}~\bibnamefont
  {Amir}}, \bibinfo {author} {\bibfnamefont {B.~P.}\ \bibnamefont {Singh}}, \
  and\ \bibinfo {author} {\bibfnamefont {S.~G.}\ \bibnamefont {Ghosh}},\ }\href
  {\doibase 10.1140/epjc/s10052-018-5872-3} {\bibfield  {journal} {\bibinfo
  {journal} {Eur. Phys. J. C}\ }\textbf {\bibinfo {volume} {78}},\ \bibinfo
  {pages} {399} (\bibinfo {year} {2018})},\ \Eprint
  {http://arxiv.org/abs/1707.09521} {arXiv:1707.09521 [gr-qc]} \BibitemShut
  {NoStop}%
\bibitem [{\citenamefont {Bambi}\ and\ \citenamefont
  {Yoshida}(2010)}]{Bambi:2010hf}%
  \BibitemOpen
  \bibfield  {author} {\bibinfo {author} {\bibfnamefont {C.}~\bibnamefont
  {Bambi}}\ and\ \bibinfo {author} {\bibfnamefont {N.}~\bibnamefont
  {Yoshida}},\ }\href {\doibase 10.1088/0264-9381/27/20/205006} {\bibfield
  {journal} {\bibinfo  {journal} {Class. Quant. Grav.}\ }\textbf {\bibinfo
  {volume} {27}},\ \bibinfo {pages} {205006} (\bibinfo {year} {2010})},\
  \Eprint {http://arxiv.org/abs/1004.3149} {arXiv:1004.3149 [gr-qc]}
  \BibitemShut {NoStop}%
\bibitem [{\citenamefont {Konoplya}\ \emph {et~al.}(2020)\citenamefont
  {Konoplya}, \citenamefont {Pappas},\ and\ \citenamefont
  {Zhidenko}}]{Konoplya:2019fpy}%
  \BibitemOpen
  \bibfield  {author} {\bibinfo {author} {\bibfnamefont {R.~A.}\ \bibnamefont
  {Konoplya}}, \bibinfo {author} {\bibfnamefont {T.}~\bibnamefont {Pappas}}, \
  and\ \bibinfo {author} {\bibfnamefont {A.}~\bibnamefont {Zhidenko}},\ }\href
  {\doibase 10.1103/PhysRevD.101.044054} {\bibfield  {journal} {\bibinfo
  {journal} {Phys. Rev. D}\ }\textbf {\bibinfo {volume} {101}},\ \bibinfo
  {pages} {044054} (\bibinfo {year} {2020})},\ \Eprint
  {http://arxiv.org/abs/1907.10112} {arXiv:1907.10112 [gr-qc]} \BibitemShut
  {NoStop}%
\bibitem [{\citenamefont {Jusufi}\ \emph {et~al.}(2020)\citenamefont {Jusufi},
  \citenamefont {Jamil},\ and\ \citenamefont {Zhu}}]{Jusufi:2020cpn}%
  \BibitemOpen
  \bibfield  {author} {\bibinfo {author} {\bibfnamefont {K.}~\bibnamefont
  {Jusufi}}, \bibinfo {author} {\bibfnamefont {M.}~\bibnamefont {Jamil}}, \
  and\ \bibinfo {author} {\bibfnamefont {T.}~\bibnamefont {Zhu}},\ }\href
  {\doibase 10.1140/epjc/s10052-020-7899-5} {\bibfield  {journal} {\bibinfo
  {journal} {Eur. Phys. J. C}\ }\textbf {\bibinfo {volume} {80}},\ \bibinfo
  {pages} {354} (\bibinfo {year} {2020})},\ \Eprint
  {http://arxiv.org/abs/2005.05299} {arXiv:2005.05299 [gr-qc]} \BibitemShut
  {NoStop}%
\bibitem [{\citenamefont {Wang}\ \emph {et~al.}(2020)\citenamefont {Wang},
  \citenamefont {Chen}, \citenamefont {Wang},\ and\ \citenamefont
  {Jing}}]{Wang:2019tjc}%
  \BibitemOpen
  \bibfield  {author} {\bibinfo {author} {\bibfnamefont {M.}~\bibnamefont
  {Wang}}, \bibinfo {author} {\bibfnamefont {S.}~\bibnamefont {Chen}}, \bibinfo
  {author} {\bibfnamefont {J.}~\bibnamefont {Wang}}, \ and\ \bibinfo {author}
  {\bibfnamefont {J.}~\bibnamefont {Jing}},\ }\href {\doibase
  10.1140/epjc/s10052-020-7641-3} {\bibfield  {journal} {\bibinfo  {journal}
  {Eur. Phys. J. C}\ }\textbf {\bibinfo {volume} {80}},\ \bibinfo {pages} {110}
  (\bibinfo {year} {2020})},\ \Eprint {http://arxiv.org/abs/1904.12423}
  {arXiv:1904.12423 [gr-qc]} \BibitemShut {NoStop}%
\bibitem [{\citenamefont {Zeng}\ \emph {et~al.}(2020)\citenamefont {Zeng},
  \citenamefont {Zhang},\ and\ \citenamefont {Zhang}}]{Zeng:2020dco}%
  \BibitemOpen
  \bibfield  {author} {\bibinfo {author} {\bibfnamefont {X.-X.}\ \bibnamefont
  {Zeng}}, \bibinfo {author} {\bibfnamefont {H.-Q.}\ \bibnamefont {Zhang}}, \
  and\ \bibinfo {author} {\bibfnamefont {H.}~\bibnamefont {Zhang}},\
  }\href@noop {} {\  (\bibinfo {year} {2020})},\ \Eprint
  {http://arxiv.org/abs/2004.12074} {arXiv:2004.12074 [gr-qc]} \BibitemShut
  {NoStop}%
\bibitem [{\citenamefont {Wei}\ \emph {et~al.}(2019{\natexlab{a}})\citenamefont
  {Wei}, \citenamefont {Zou}, \citenamefont {Liu},\ and\ \citenamefont
  {Mann}}]{Wei:2019pjf}%
  \BibitemOpen
  \bibfield  {author} {\bibinfo {author} {\bibfnamefont {S.-W.}\ \bibnamefont
  {Wei}}, \bibinfo {author} {\bibfnamefont {Y.-C.}\ \bibnamefont {Zou}},
  \bibinfo {author} {\bibfnamefont {Y.-X.}\ \bibnamefont {Liu}}, \ and\
  \bibinfo {author} {\bibfnamefont {R.~B.}\ \bibnamefont {Mann}},\ }\href
  {\doibase 10.1088/1475-7516/2019/08/030} {\bibfield  {journal} {\bibinfo
  {journal} {JCAP}\ }\textbf {\bibinfo {volume} {1908}},\ \bibinfo {pages}
  {030} (\bibinfo {year} {2019}{\natexlab{a}})},\ \Eprint
  {http://arxiv.org/abs/1904.07710} {arXiv:1904.07710 [gr-qc]} \BibitemShut
  {NoStop}%
\bibitem [{\citenamefont {Ohgami}\ and\ \citenamefont
  {Sakai}(2015)}]{Ohgami:2015nra}%
  \BibitemOpen
  \bibfield  {author} {\bibinfo {author} {\bibfnamefont {T.}~\bibnamefont
  {Ohgami}}\ and\ \bibinfo {author} {\bibfnamefont {N.}~\bibnamefont {Sakai}},\
  }\href {\doibase 10.1103/PhysRevD.91.124020} {\bibfield  {journal} {\bibinfo
  {journal} {Phys. Rev.}\ }\textbf {\bibinfo {volume} {D91}},\ \bibinfo {pages}
  {124020} (\bibinfo {year} {2015})},\ \Eprint
  {http://arxiv.org/abs/1704.07065} {arXiv:1704.07065 [gr-qc]} \BibitemShut
  {NoStop}%
\bibitem [{\citenamefont {Nedkova}\ \emph {et~al.}(2013)\citenamefont
  {Nedkova}, \citenamefont {Tinchev},\ and\ \citenamefont
  {Yazadjiev}}]{Nedkova:2013msa}%
  \BibitemOpen
  \bibfield  {author} {\bibinfo {author} {\bibfnamefont {P.~G.}\ \bibnamefont
  {Nedkova}}, \bibinfo {author} {\bibfnamefont {V.~K.}\ \bibnamefont
  {Tinchev}}, \ and\ \bibinfo {author} {\bibfnamefont {S.~S.}\ \bibnamefont
  {Yazadjiev}},\ }\href {\doibase 10.1103/PhysRevD.88.124019} {\bibfield
  {journal} {\bibinfo  {journal} {Phys. Rev.}\ }\textbf {\bibinfo {volume}
  {D88}},\ \bibinfo {pages} {124019} (\bibinfo {year} {2013})},\ \Eprint
  {http://arxiv.org/abs/1307.7647} {arXiv:1307.7647 [gr-qc]} \BibitemShut
  {NoStop}%
\bibitem [{\citenamefont {Shaikh}(2018)}]{Shaikh:2018kfv}%
  \BibitemOpen
  \bibfield  {author} {\bibinfo {author} {\bibfnamefont {R.}~\bibnamefont
  {Shaikh}},\ }\href {\doibase 10.1103/PhysRevD.98.024044} {\bibfield
  {journal} {\bibinfo  {journal} {Phys. Rev.}\ }\textbf {\bibinfo {volume}
  {D98}},\ \bibinfo {pages} {024044} (\bibinfo {year} {2018})},\ \Eprint
  {http://arxiv.org/abs/1803.11422} {arXiv:1803.11422 [gr-qc]} \BibitemShut
  {NoStop}%
\bibitem [{\citenamefont {Amir}\ \emph
  {et~al.}(2019{\natexlab{a}})\citenamefont {Amir}, \citenamefont {Banerjee},\
  and\ \citenamefont {Maharaj}}]{Amir:2018szm}%
  \BibitemOpen
  \bibfield  {author} {\bibinfo {author} {\bibfnamefont {M.}~\bibnamefont
  {Amir}}, \bibinfo {author} {\bibfnamefont {A.}~\bibnamefont {Banerjee}}, \
  and\ \bibinfo {author} {\bibfnamefont {S.~D.}\ \bibnamefont {Maharaj}},\
  }\href {\doibase 10.1016/j.aop.2018.11.004} {\bibfield  {journal} {\bibinfo
  {journal} {Annals Phys.}\ }\textbf {\bibinfo {volume} {400}},\ \bibinfo
  {pages} {198} (\bibinfo {year} {2019}{\natexlab{a}})},\ \Eprint
  {http://arxiv.org/abs/1805.12435} {arXiv:1805.12435 [gr-qc]} \BibitemShut
  {NoStop}%
\bibitem [{\citenamefont {Amir}\ \emph
  {et~al.}(2019{\natexlab{b}})\citenamefont {Amir}, \citenamefont {Jusufi},
  \citenamefont {Banerjee},\ and\ \citenamefont {Hansraj}}]{Amir:2018pcu}%
  \BibitemOpen
  \bibfield  {author} {\bibinfo {author} {\bibfnamefont {M.}~\bibnamefont
  {Amir}}, \bibinfo {author} {\bibfnamefont {K.}~\bibnamefont {Jusufi}},
  \bibinfo {author} {\bibfnamefont {A.}~\bibnamefont {Banerjee}}, \ and\
  \bibinfo {author} {\bibfnamefont {S.}~\bibnamefont {Hansraj}},\ }\href
  {\doibase 10.1088/1361-6382/ab42be} {\bibfield  {journal} {\bibinfo
  {journal} {Class. Quant. Grav.}\ }\textbf {\bibinfo {volume} {36}},\ \bibinfo
  {pages} {215007} (\bibinfo {year} {2019}{\natexlab{b}})},\ \Eprint
  {http://arxiv.org/abs/1806.07782} {arXiv:1806.07782 [gr-qc]} \BibitemShut
  {NoStop}%
\bibitem [{\citenamefont {Wang}\ \emph {et~al.}(2019)\citenamefont {Wang},
  \citenamefont {Chen},\ and\ \citenamefont {Jing}}]{Wang:2019skw}%
  \BibitemOpen
  \bibfield  {author} {\bibinfo {author} {\bibfnamefont {M.}~\bibnamefont
  {Wang}}, \bibinfo {author} {\bibfnamefont {S.}~\bibnamefont {Chen}}, \ and\
  \bibinfo {author} {\bibfnamefont {J.}~\bibnamefont {Jing}},\ }\href@noop {}
  {\  (\bibinfo {year} {2019})},\ \Eprint {http://arxiv.org/abs/1908.04527}
  {arXiv:1908.04527 [gr-qc]} \BibitemShut {NoStop}%
\bibitem [{\citenamefont {Cunha}\ and\ \citenamefont
  {Herdeiro}(2018)}]{Cunha:2018acu}%
  \BibitemOpen
  \bibfield  {author} {\bibinfo {author} {\bibfnamefont {P.~V.~P.}\
  \bibnamefont {Cunha}}\ and\ \bibinfo {author} {\bibfnamefont {C.~A.~R.}\
  \bibnamefont {Herdeiro}},\ }\href {\doibase 10.1007/s10714-018-2361-9}
  {\bibfield  {journal} {\bibinfo  {journal} {Gen. Rel. Grav.}\ }\textbf
  {\bibinfo {volume} {50}},\ \bibinfo {pages} {42} (\bibinfo {year} {2018})},\
  \Eprint {http://arxiv.org/abs/1801.00860} {arXiv:1801.00860 [gr-qc]}
  \BibitemShut {NoStop}%
\bibitem [{\citenamefont {Kubiznak}\ and\ \citenamefont
  {Mann}(2012)}]{Kubiznak:2012wp}%
  \BibitemOpen
  \bibfield  {author} {\bibinfo {author} {\bibfnamefont {D.}~\bibnamefont
  {Kubiznak}}\ and\ \bibinfo {author} {\bibfnamefont {R.~B.}\ \bibnamefont
  {Mann}},\ }\href {\doibase 10.1007/JHEP07(2012)033} {\bibfield  {journal}
  {\bibinfo  {journal} {JHEP}\ }\textbf {\bibinfo {volume} {07}},\ \bibinfo
  {pages} {033} (\bibinfo {year} {2012})},\ \Eprint
  {http://arxiv.org/abs/1205.0559} {arXiv:1205.0559 [hep-th]} \BibitemShut
  {NoStop}%
\bibitem [{\citenamefont {Wei}\ \emph {et~al.}(2019{\natexlab{b}})\citenamefont
  {Wei}, \citenamefont {Liu},\ and\ \citenamefont {Mann}}]{Wei:2019uqg}%
  \BibitemOpen
  \bibfield  {author} {\bibinfo {author} {\bibfnamefont {S.-W.}\ \bibnamefont
  {Wei}}, \bibinfo {author} {\bibfnamefont {Y.-X.}\ \bibnamefont {Liu}}, \ and\
  \bibinfo {author} {\bibfnamefont {R.~B.}\ \bibnamefont {Mann}},\ }\href
  {\doibase 10.1103/PhysRevLett.123.071103} {\bibfield  {journal} {\bibinfo
  {journal} {Phys. Rev. Lett.}\ }\textbf {\bibinfo {volume} {123}},\ \bibinfo
  {pages} {071103} (\bibinfo {year} {2019}{\natexlab{b}})},\ \Eprint
  {http://arxiv.org/abs/1906.10840} {arXiv:1906.10840 [gr-qc]} \BibitemShut
  {NoStop}%
\bibitem [{\citenamefont {Cai}\ \emph {et~al.}(2013)\citenamefont {Cai},
  \citenamefont {Cao}, \citenamefont {Li},\ and\ \citenamefont
  {Yang}}]{Cai:2013qga}%
  \BibitemOpen
  \bibfield  {author} {\bibinfo {author} {\bibfnamefont {R.-G.}\ \bibnamefont
  {Cai}}, \bibinfo {author} {\bibfnamefont {L.-M.}\ \bibnamefont {Cao}},
  \bibinfo {author} {\bibfnamefont {L.}~\bibnamefont {Li}}, \ and\ \bibinfo
  {author} {\bibfnamefont {R.-Q.}\ \bibnamefont {Yang}},\ }\href {\doibase
  10.1007/JHEP09(2013)005} {\bibfield  {journal} {\bibinfo  {journal} {JHEP}\
  }\textbf {\bibinfo {volume} {09}},\ \bibinfo {pages} {005} (\bibinfo {year}
  {2013})},\ \Eprint {http://arxiv.org/abs/1306.6233} {arXiv:1306.6233 [gr-qc]}
  \BibitemShut {NoStop}%
\bibitem [{\citenamefont {Zhang}\ \emph {et~al.}(2018)\citenamefont {Zhang},
  \citenamefont {Wang},\ and\ \citenamefont {Liu}}]{Zhang:2018djl}%
  \BibitemOpen
  \bibfield  {author} {\bibinfo {author} {\bibfnamefont {M.}~\bibnamefont
  {Zhang}}, \bibinfo {author} {\bibfnamefont {X.-Y.}\ \bibnamefont {Wang}}, \
  and\ \bibinfo {author} {\bibfnamefont {W.-B.}\ \bibnamefont {Liu}},\ }\href
  {\doibase 10.1016/j.physletb.2018.06.061} {\bibfield  {journal} {\bibinfo
  {journal} {Phys. Lett.}\ }\textbf {\bibinfo {volume} {B783}},\ \bibinfo
  {pages} {169} (\bibinfo {year} {2018})}\BibitemShut {NoStop}%
\bibitem [{\citenamefont {Bardeen}\ \emph {et~al.}(1973)\citenamefont
  {Bardeen}, \citenamefont {Carter},\ and\ \citenamefont
  {Hawking}}]{Bardeen:1973gs}%
  \BibitemOpen
  \bibfield  {author} {\bibinfo {author} {\bibfnamefont {J.~M.}\ \bibnamefont
  {Bardeen}}, \bibinfo {author} {\bibfnamefont {B.}~\bibnamefont {Carter}}, \
  and\ \bibinfo {author} {\bibfnamefont {S.~W.}\ \bibnamefont {Hawking}},\
  }\href {\doibase 10.1007/BF01645742} {\bibfield  {journal} {\bibinfo
  {journal} {Commun. Math. Phys.}\ }\textbf {\bibinfo {volume} {31}},\ \bibinfo
  {pages} {161} (\bibinfo {year} {1973})}\BibitemShut {NoStop}%
\bibitem [{\citenamefont {Wei}\ \emph {et~al.}(2019{\natexlab{c}})\citenamefont
  {Wei}, \citenamefont {Liu},\ and\ \citenamefont {Wang}}]{Wei:2018aqm}%
  \BibitemOpen
  \bibfield  {author} {\bibinfo {author} {\bibfnamefont {S.-W.}\ \bibnamefont
  {Wei}}, \bibinfo {author} {\bibfnamefont {Y.-X.}\ \bibnamefont {Liu}}, \ and\
  \bibinfo {author} {\bibfnamefont {Y.-Q.}\ \bibnamefont {Wang}},\ }\href
  {\doibase 10.1103/PhysRevD.99.044013} {\bibfield  {journal} {\bibinfo
  {journal} {Phys. Rev.}\ }\textbf {\bibinfo {volume} {D99}},\ \bibinfo {pages}
  {044013} (\bibinfo {year} {2019}{\natexlab{c}})},\ \Eprint
  {http://arxiv.org/abs/1807.03455} {arXiv:1807.03455 [gr-qc]} \BibitemShut
  {NoStop}%
\bibitem [{\citenamefont {Han}\ \emph {et~al.}(2018)\citenamefont {Han},
  \citenamefont {Jiang}, \citenamefont {Zhang},\ and\ \citenamefont
  {Liu}}]{Han:2018ooi}%
  \BibitemOpen
  \bibfield  {author} {\bibinfo {author} {\bibfnamefont {S.-Z.}\ \bibnamefont
  {Han}}, \bibinfo {author} {\bibfnamefont {J.}~\bibnamefont {Jiang}}, \bibinfo
  {author} {\bibfnamefont {M.}~\bibnamefont {Zhang}}, \ and\ \bibinfo {author}
  {\bibfnamefont {W.-B.}\ \bibnamefont {Liu}},\ }\href@noop {} {\  (\bibinfo
  {year} {2018})},\ \Eprint {http://arxiv.org/abs/1812.11862} {arXiv:1812.11862
  [gr-qc]} \BibitemShut {NoStop}%
\bibitem [{\citenamefont {Zhang}\ \emph {et~al.}(2019)\citenamefont {Zhang},
  \citenamefont {Han}, \citenamefont {Jiang},\ and\ \citenamefont
  {Liu}}]{Zhang:2019tzi}%
  \BibitemOpen
  \bibfield  {author} {\bibinfo {author} {\bibfnamefont {M.}~\bibnamefont
  {Zhang}}, \bibinfo {author} {\bibfnamefont {S.-Z.}\ \bibnamefont {Han}},
  \bibinfo {author} {\bibfnamefont {J.}~\bibnamefont {Jiang}}, \ and\ \bibinfo
  {author} {\bibfnamefont {W.-B.}\ \bibnamefont {Liu}},\ }\href {\doibase
  10.1103/PhysRevD.99.065016} {\bibfield  {journal} {\bibinfo  {journal} {Phys.
  Rev.}\ }\textbf {\bibinfo {volume} {D99}},\ \bibinfo {pages} {065016}
  (\bibinfo {year} {2019})},\ \Eprint {http://arxiv.org/abs/1903.08293}
  {arXiv:1903.08293 [hep-th]} \BibitemShut {NoStop}%
\bibitem [{\citenamefont {Xu}\ \emph {et~al.}(2019)\citenamefont {Xu},
  \citenamefont {Wang}, \citenamefont {Liu},\ and\ \citenamefont
  {Wei}}]{Xu:2019yub}%
  \BibitemOpen
  \bibfield  {author} {\bibinfo {author} {\bibfnamefont {Y.-M.}\ \bibnamefont
  {Xu}}, \bibinfo {author} {\bibfnamefont {H.-M.}\ \bibnamefont {Wang}},
  \bibinfo {author} {\bibfnamefont {Y.-X.}\ \bibnamefont {Liu}}, \ and\
  \bibinfo {author} {\bibfnamefont {S.-W.}\ \bibnamefont {Wei}},\ }\href
  {\doibase 10.1103/PhysRevD.100.104044} {\bibfield  {journal} {\bibinfo
  {journal} {Phys. Rev. D}\ }\textbf {\bibinfo {volume} {100}},\ \bibinfo
  {pages} {104044} (\bibinfo {year} {2019})},\ \Eprint
  {http://arxiv.org/abs/1906.03334} {arXiv:1906.03334 [gr-qc]} \BibitemShut
  {NoStop}%
\bibitem [{\citenamefont {Li}\ \emph {et~al.}(2020{\natexlab{a}})\citenamefont
  {Li}, \citenamefont {Chen},\ and\ \citenamefont {Zhang}}]{Li:2019dai}%
  \BibitemOpen
  \bibfield  {author} {\bibinfo {author} {\bibfnamefont {H.}~\bibnamefont
  {Li}}, \bibinfo {author} {\bibfnamefont {Y.}~\bibnamefont {Chen}}, \ and\
  \bibinfo {author} {\bibfnamefont {S.-J.}\ \bibnamefont {Zhang}},\ }\href
  {\doibase 10.1016/j.nuclphysb.2020.114975} {\bibfield  {journal} {\bibinfo
  {journal} {Nucl. Phys. B}\ }\textbf {\bibinfo {volume} {954}},\ \bibinfo
  {pages} {114975} (\bibinfo {year} {2020}{\natexlab{a}})},\ \Eprint
  {http://arxiv.org/abs/1908.09570} {arXiv:1908.09570 [hep-th]} \BibitemShut
  {NoStop}%
\bibitem [{\citenamefont {Li}\ \emph {et~al.}(2020{\natexlab{b}})\citenamefont
  {Li}, \citenamefont {Guo},\ and\ \citenamefont {Chen}}]{Li:2020drn}%
  \BibitemOpen
  \bibfield  {author} {\bibinfo {author} {\bibfnamefont {P.-C.}\ \bibnamefont
  {Li}}, \bibinfo {author} {\bibfnamefont {M.}~\bibnamefont {Guo}}, \ and\
  \bibinfo {author} {\bibfnamefont {B.}~\bibnamefont {Chen}},\ }\href {\doibase
  10.1103/PhysRevD.101.084041} {\bibfield  {journal} {\bibinfo  {journal}
  {Phys. Rev. D}\ }\textbf {\bibinfo {volume} {101}},\ \bibinfo {pages}
  {084041} (\bibinfo {year} {2020}{\natexlab{b}})},\ \Eprint
  {http://arxiv.org/abs/2001.04231} {arXiv:2001.04231 [gr-qc]} \BibitemShut
  {NoStop}%
\bibitem [{\citenamefont {Stuchlík}\ \emph {et~al.}(2018)\citenamefont
  {Stuchlík}, \citenamefont {Charbulák},\ and\ \citenamefont
  {Schee}}]{Stuchlik:2018qyz}%
  \BibitemOpen
  \bibfield  {author} {\bibinfo {author} {\bibfnamefont {Z.}~\bibnamefont
  {Stuchlík}}, \bibinfo {author} {\bibfnamefont {D.}~\bibnamefont
  {Charbulák}}, \ and\ \bibinfo {author} {\bibfnamefont {J.}~\bibnamefont
  {Schee}},\ }\href {\doibase 10.1140/epjc/s10052-018-5578-6} {\bibfield
  {journal} {\bibinfo  {journal} {Eur. Phys. J.}\ }\textbf {\bibinfo {volume}
  {C78}},\ \bibinfo {pages} {180} (\bibinfo {year} {2018})},\ \Eprint
  {http://arxiv.org/abs/1811.00072} {arXiv:1811.00072 [gr-qc]} \BibitemShut
  {NoStop}%
\bibitem [{\citenamefont {Cunha}\ \emph {et~al.}(2016)\citenamefont {Cunha},
  \citenamefont {Herdeiro}, \citenamefont {Radu},\ and\ \citenamefont
  {Runarsson}}]{Cunha:2016bpi}%
  \BibitemOpen
  \bibfield  {author} {\bibinfo {author} {\bibfnamefont {P.~V.~P.}\
  \bibnamefont {Cunha}}, \bibinfo {author} {\bibfnamefont {C.~A.~R.}\
  \bibnamefont {Herdeiro}}, \bibinfo {author} {\bibfnamefont {E.}~\bibnamefont
  {Radu}}, \ and\ \bibinfo {author} {\bibfnamefont {H.~F.}\ \bibnamefont
  {Runarsson}},\ }\href {\doibase 10.1142/S0218271816410212} {\bibfield
  {journal} {\bibinfo  {journal} {Int. J. Mod. Phys.}\ }\textbf {\bibinfo
  {volume} {D25}},\ \bibinfo {pages} {1641021} (\bibinfo {year} {2016})},\
  \Eprint {http://arxiv.org/abs/1605.08293} {arXiv:1605.08293 [gr-qc]}
  \BibitemShut {NoStop}%
\bibitem [{\citenamefont {Hioki}\ and\ \citenamefont
  {Maeda}(2009)}]{Hioki:2009na}%
  \BibitemOpen
  \bibfield  {author} {\bibinfo {author} {\bibfnamefont {K.}~\bibnamefont
  {Hioki}}\ and\ \bibinfo {author} {\bibfnamefont {K.-i.}\ \bibnamefont
  {Maeda}},\ }\href {\doibase 10.1103/PhysRevD.80.024042} {\bibfield  {journal}
  {\bibinfo  {journal} {Phys. Rev.}\ }\textbf {\bibinfo {volume} {D80}},\
  \bibinfo {pages} {024042} (\bibinfo {year} {2009})},\ \Eprint
  {http://arxiv.org/abs/0904.3575} {arXiv:0904.3575 [astro-ph.HE]} \BibitemShut
  {NoStop}%
\bibitem [{\citenamefont {Feng}\ and\ \citenamefont {Lu}(2020)}]{Feng:2019zzn}%
  \BibitemOpen
  \bibfield  {author} {\bibinfo {author} {\bibfnamefont {X.-H.}\ \bibnamefont
  {Feng}}\ and\ \bibinfo {author} {\bibfnamefont {H.}~\bibnamefont {Lu}},\
  }\href {\doibase 10.1140/epjc/s10052-020-8119-z} {\bibfield  {journal}
  {\bibinfo  {journal} {Eur. Phys. J. C}\ }\textbf {\bibinfo {volume} {80}},\
  \bibinfo {pages} {551} (\bibinfo {year} {2020})},\ \Eprint
  {http://arxiv.org/abs/1911.12368} {arXiv:1911.12368 [gr-qc]} \BibitemShut
  {NoStop}%
\bibitem [{\citenamefont {Caldarelli}\ \emph {et~al.}(2000)\citenamefont
  {Caldarelli}, \citenamefont {Cognola},\ and\ \citenamefont
  {Klemm}}]{Caldarelli:1999xj}%
  \BibitemOpen
  \bibfield  {author} {\bibinfo {author} {\bibfnamefont {M.~M.}\ \bibnamefont
  {Caldarelli}}, \bibinfo {author} {\bibfnamefont {G.}~\bibnamefont {Cognola}},
  \ and\ \bibinfo {author} {\bibfnamefont {D.}~\bibnamefont {Klemm}},\ }\href
  {\doibase 10.1088/0264-9381/17/2/310} {\bibfield  {journal} {\bibinfo
  {journal} {Class. Quant. Grav.}\ }\textbf {\bibinfo {volume} {17}},\ \bibinfo
  {pages} {399} (\bibinfo {year} {2000})},\ \Eprint
  {http://arxiv.org/abs/hep-th/9908022} {arXiv:hep-th/9908022 [hep-th]}
  \BibitemShut {NoStop}%
\bibitem [{\citenamefont {Gibbons}\ \emph {et~al.}(2005)\citenamefont
  {Gibbons}, \citenamefont {Perry},\ and\ \citenamefont
  {Pope}}]{Gibbons:2004ai}%
  \BibitemOpen
  \bibfield  {author} {\bibinfo {author} {\bibfnamefont {G.~W.}\ \bibnamefont
  {Gibbons}}, \bibinfo {author} {\bibfnamefont {M.~J.}\ \bibnamefont {Perry}},
  \ and\ \bibinfo {author} {\bibfnamefont {C.~N.}\ \bibnamefont {Pope}},\
  }\href {\doibase 10.1088/0264-9381/22/9/002} {\bibfield  {journal} {\bibinfo
  {journal} {Class. Quant. Grav.}\ }\textbf {\bibinfo {volume} {22}},\ \bibinfo
  {pages} {1503} (\bibinfo {year} {2005})},\ \Eprint
  {http://arxiv.org/abs/hep-th/0408217} {arXiv:hep-th/0408217 [hep-th]}
  \BibitemShut {NoStop}%
\bibitem [{\citenamefont {Cvetic}\ \emph {et~al.}(2011)\citenamefont {Cvetic},
  \citenamefont {Gibbons}, \citenamefont {Kubiznak},\ and\ \citenamefont
  {Pope}}]{Cvetic:2010jb}%
  \BibitemOpen
  \bibfield  {author} {\bibinfo {author} {\bibfnamefont {M.}~\bibnamefont
  {Cvetic}}, \bibinfo {author} {\bibfnamefont {G.}~\bibnamefont {Gibbons}},
  \bibinfo {author} {\bibfnamefont {D.}~\bibnamefont {Kubiznak}}, \ and\
  \bibinfo {author} {\bibfnamefont {C.}~\bibnamefont {Pope}},\ }\href {\doibase
  10.1103/PhysRevD.84.024037} {\bibfield  {journal} {\bibinfo  {journal} {Phys.
  Rev. D}\ }\textbf {\bibinfo {volume} {84}},\ \bibinfo {pages} {024037}
  (\bibinfo {year} {2011})},\ \Eprint {http://arxiv.org/abs/1012.2888}
  {arXiv:1012.2888 [hep-th]} \BibitemShut {NoStop}%
\bibitem [{\citenamefont {Gunasekaran}\ \emph {et~al.}(2012)\citenamefont
  {Gunasekaran}, \citenamefont {Mann},\ and\ \citenamefont
  {Kubiznak}}]{Gunasekaran:2012dq}%
  \BibitemOpen
  \bibfield  {author} {\bibinfo {author} {\bibfnamefont {S.}~\bibnamefont
  {Gunasekaran}}, \bibinfo {author} {\bibfnamefont {R.~B.}\ \bibnamefont
  {Mann}}, \ and\ \bibinfo {author} {\bibfnamefont {D.}~\bibnamefont
  {Kubiznak}},\ }\href {\doibase 10.1007/JHEP11(2012)110} {\bibfield  {journal}
  {\bibinfo  {journal} {JHEP}\ }\textbf {\bibinfo {volume} {11}},\ \bibinfo
  {pages} {110} (\bibinfo {year} {2012})},\ \Eprint
  {http://arxiv.org/abs/1208.6251} {arXiv:1208.6251 [hep-th]} \BibitemShut
  {NoStop}%
\bibitem [{\citenamefont {Dolan}\ \emph {et~al.}(2013)\citenamefont {Dolan},
  \citenamefont {Kastor}, \citenamefont {Kubiznak}, \citenamefont {Mann},\ and\
  \citenamefont {Traschen}}]{Dolan:2013ft}%
  \BibitemOpen
  \bibfield  {author} {\bibinfo {author} {\bibfnamefont {B.~P.}\ \bibnamefont
  {Dolan}}, \bibinfo {author} {\bibfnamefont {D.}~\bibnamefont {Kastor}},
  \bibinfo {author} {\bibfnamefont {D.}~\bibnamefont {Kubiznak}}, \bibinfo
  {author} {\bibfnamefont {R.~B.}\ \bibnamefont {Mann}}, \ and\ \bibinfo
  {author} {\bibfnamefont {J.}~\bibnamefont {Traschen}},\ }\href {\doibase
  10.1103/PhysRevD.87.104017} {\bibfield  {journal} {\bibinfo  {journal} {Phys.
  Rev. D}\ }\textbf {\bibinfo {volume} {87}},\ \bibinfo {pages} {104017}
  (\bibinfo {year} {2013})},\ \Eprint {http://arxiv.org/abs/1301.5926}
  {arXiv:1301.5926 [hep-th]} \BibitemShut {NoStop}%
\end{thebibliography}
\end{document}